\documentclass{emulateapj-rtx4}
\usepackage{amsmath}
\usepackage{float}
\makeatletter
\def\gsim{\compoundrel>\over\sim}
\def\lsim{\compoundrel<\over\sim}
\def\compoundrel#1\over#2{\mathpalette\compoundreL{{#1}\over{#2}}}
\def\compoundreL#1#2{\compoundREL#1#2}
\def\compoundREL#1#2\over#3{\mathrel
      {\vcenter{\hbox{$\m@th\buildrel{#1#2}\over{#1#3}$}}}}
\usepackage{subfigure}

\slugcomment{}

\shorttitle{Polarization and Spectral Variability in M87}
\shortauthors{Perlman et al.}

\begin{document}

\setcounter{page}{01}

\def\lsim{\lower.5ex\hbox{$\; \buildrel < \over \sim \;$}}
\def\gsim{\lower.5ex\hbox{$\; \buildrel > \over \sim \;$}}
 

\title{Optical Polarization and Spectral Variability in the M87 Jet}


\author{Eric S. Perlman \altaffilmark{1}, Steven C. Adams\altaffilmark{2,3},
Mihai Cara\altaffilmark{1,4}, Matthew Bourque\altaffilmark{1,3}, 
D. E. Harris\altaffilmark{5}, Juan P. Madrid\altaffilmark{6},  
Raymond C. Simons\altaffilmark{1},
Eric Clausen-Brown\altaffilmark{7}, C. C. Cheung\altaffilmark{8},
Lukasz Stawarz\altaffilmark{9,10}, Markos Georganopoulos\altaffilmark{11},
William B. Sparks\altaffilmark{12}, John A. Biretta\altaffilmark{12}}

\altaffiltext{1}{Department of Physics and Space Sciences, 150 W. University
Blvd., Florida Institute of Technology, Melbourne, FL 32901}

\altaffiltext{2}{Department of Physics and Astronomy, University of Georgia,
Athens, GA 30605}

\altaffiltext{3}{Southeastern Association for Research in Astronomy (SARA)
NSF-REU Summer Intern}

\altaffiltext{4}{Current address:  Physics Department, Case Western Reserve
University, 10900 Euclid Ave., Cleveland, OH  44106-7079}

\altaffiltext{5}{Harvard-Smithsonian Center for Astrophysics, 60 Garden Street,
Cambridge, MA  02138}

\altaffiltext{6}{Center for Astrophysics and Supercomputing, 
Swinburne University of Technology, Hawthorn, VIC 3122, Australia}

\altaffiltext{7}{Department of Physics, Purdue University, West Lafayette, IN
47907}

\altaffiltext{8}{National Research Council Research Associate, National Academy
of Sciences, Washington, DC 20001, resident at Naval Research Laboratory, 
Washington, DC  20375}

\altaffiltext{9}{Institute of Space and Astronautical Science, JAXA, 3-1-1
Yoshinodai, Chuo-ku, Sagamihara, Kanagawa 252-5210, Japan}

\altaffiltext{10}{Astronomical Observatory, Jagiellonian University, ul. Orla
171, 30-244 Krak\'ow, Poland} 

\altaffiltext{11}{Department of Physics, University of Maryland - Baltimore
County, 1000 Hilltop Circle, Baltimore, MD  21250} 

\altaffiltext{12}{Space Telescope Science Institute, 3700 San Martin Drive, Baltimore,
MD, 21218}

\email{eperlman@fit.edu}


\newcommand{\vdag}{(v)^\dagger}


\begin{abstract}

During the last decade, M87's jet has been the site of an 
extraordinary variability event,  with one knot (HST-1) increasing by over 
a factor 100 in brightness.  
Variability was also seen on timescales of months in
the nuclear flux.  
Here we discuss the optical-UV polarization and spectral variability of these
components, which show vastly different behavior.  HST-1 shows a 
highly significant correlation between flux and polarization,
with $P$ increasing from  $\sim 20\%$ at minimum to
$>40\%$ at  maximum, while the orientation  of its
electric vector stayed constant.  HST-1's optical-UV spectrum is 
very hard
($\alpha_{UV-O}\sim0.5$, $F_\nu\propto\nu^{-\alpha}$), and displays 
``hard lags" during epochs 2004.9-2005.5, 
including the peak of the flare, with soft lags at
later epochs.  
We interpret the behavior of HST-1 as enhanced particle acceleration in
a shock, with cooling from both particle aging and  
the relaxation of the compression.  We set 2$\sigma$ upper limits 
of $0.5 \delta$ parsecs and 1.02$c$ on the size and advance speed of 
the flaring region.
The slight deviation of the electric vector orientation
from the jet PA, makes it likely that 
on smaller scales the 
flaring region has either a double or twisted structure.
By contrast, the nucleus displays much more rapid
variability, with a highly variable electric vector orientation and 
'looping' in the $(I,P)$ plane.   
The nucleus has a much steeper spectrum  
($\alpha_{UV-O} \sim 1.5$) but does not show 
UV-optical spectral variability.
Its behavior can be interpreted as either a helical distortion
to a steady jet or a shock propagating through a helical jet. 

\end{abstract}




\section{Introduction}

M87's jet was one of the first manifestations observed of the active galactic
nucleus (AGN) phenomenon (Curtis 1918), and has been the target of myriad
observations due to its brightness and also proximity ($d=16$ Mpc, Tonry 1991).
During the last decade, M87's jet has been the site of an extraordinary
variability event, with knot HST-1  increasing in optical/UV brightness by a
factor of more than 100 between  2000 and its peak in 2005.  The flare in knot
HST-1 has been the target of several monitoring efforts using the {\it Hubble
Space Telescope} (hereafter {\it HST}), {\it Chandra X-ray Observatory}, {\it
VLA} and other telescopes.  Previous papers from this project include Paper I
reporting our first results \citep{har03}, Paper II which focused on the HST data
\citep{per03}, Paper III which was mainly on the X-ray lightcurve of  HST-1 which
delineated the massive 2005 flare (Harris et al. 2006), Paper IV which focussed
on the VLBA results, showing superluminal proper motions in HST-1  (Cheung et al.
2007), and Paper V (Harris et al. 2009)  which focused on a more detailed
analysis of the variability timescales  of HST-1 and the nucleus.  Madrid (2009)
has also  added a complete analysis of the UV light-curve of HST-1 and the
nucleus between 2000-2006.

In this paper (VI of the series), we discuss two additional aspects of the 
monitoring campaign, namely the evolution of the polarization and spectral 
index in the optical-UV.  We concentrate on the nucleus and HST-1, 
as they are the main variable components in the jet.  A future
paper will combine these observations to produce a new polarization map
of the entire jet and discuss any changes over the decade between the data 
of Perlman
et al. (1999) and this paper.  Section 2 provides a detailed explanation
of the observations and data reduction process.
Then, in Section 3, we will discuss the techniques used in
reducing the data.  In section 4, we will discuss results.
Finally, in Section 5, we discuss implications for jet variability models
and close in Section 6 with a summary.

\section{Observations}

As has been well documented (e.g., Paper I, Waters \& Zepf 2005), 
the flare of knot HST-1 began sometime during 
2000.  While the jet of M87 was a regular target for {\it HST} and {\it Chandra}
almost from the start, intensive monitoring by these telescopes began in 
2002 (Papers I, II, III, Madrid 2009).
Here we review these observations, concentrating on the optical polarimetric
part of the campaign.  

Optical polarimetry was obtained on a somewhat 
different schedule than the UV imaging discussed in Madrid (2009).  
Of the eighteen observations obtained 
between 2002 December and 2007 November, fourteen occurred 
on the same schedule as the 
imaging observations during November 2004-December 2006.  The other four
observations were at roughly yearly intervals before and after this period.
Polarimetry was done in two bands, F330W and F606W, 
with F606W observations being done much more often. 
Table 1 details the scheduling of these observations, along with other
information about the nucleus and HST-1 which will be described later.

The High-Resolution Channel (HRC) of the
Advanced Camera for Surveys (ACS) was used to obtain the polarimetry 
data for 17 of the 18 epochs. 
The ACS HRC is a single-chip CCD camera, with a plate scale of 
0.028$\times$0.025 arcsec/pixel, corresponding to a field of view of about
28$\times$ 25 arcseconds and yielding diffraction limited resolution of 
$\approx$0".06 for the F606W observations, and a 
Nyquist-limited resolution of $\approx$ 0".05 for the F330W observations
(see Maybhate et al. 2010). On ACS the polarizing set is comprised of either 
the POL0V, POL60V and POL120V filters (used for F606W) 
or the POL0UV, POL60UV, and POL120UV filters (used for F330W).
As the three polarization
filters can be selected individually \cite{May10},
no change in position or chip was necessary between them.  All observations
were CR-SPLIT to  mitigate the effects of cosmic rays, but dithering was 
typically not done.

For the final epoch, which occurred
after ACS went offline due to an electrical short, 
the Wide Field Planetary Camera 2 (WFPC2) was used for polarimetry.  
On WFPC2 the polarizing filter
set is comprised of the POLQ quad, which has filters at angles of 0, 45 and 
90 degreees.  The WFPC2 is a 
chevron-shaped, four-CCD camera, with three wide-field chips (WF2, WF3, and WF4)
and a fine-resolution one (PC1).  Its setup is less flexible for polarimetry,
as the POLQ quad can only be rotated through 51 degrees (Biretta \& McMaster
1997).  For this reason, we used the three WF chips.  These chips have a 
plate scale of 0.09965
arcsec/pixel, leading to a final resolution $\sim 0.2"$ set by the 
Nyquist theorem.

\section{Data Reduction}

We obtained the data for these observations from the {\it HST} archive.  All
data were recalibrated with updated flat field files and image distortion
correction (IDC) tables, obtained from from the STScI
Calibration Database System. Standard techniques were used to
recalibrate the data. These methods are described in detail in the instrument
handbooks for ACS and WFPC2 \citep{May10,mob02}. CTE corrections were computed
using the data found in the ACS Instrument Handbook \citep{May10}.

After the data were retrieved, MULTIDRIZZLE was used to
combine and cosmic-ray reject
the images. The alignment of images was refined, assuming the positions of the 
core and HST-1 as canonical and using TWEAKSHIFTS in IRAF to shift the images to
a common frame of reference and correct for geometric distortions using the 
models in the IDC. This was done because experience with the polarizing filters
on both ACS and WFPC-2 has shown that there can be small
irregularities in the filters 
that can cause sources that are far away from one's region of
interest to shift apparently as compared to other sources in the field. 
The procedure was also checked by using only the nucleus as canonical, 
thus allowing for the possibility of motions in HST-1. 
Unfortunately with the short exposure times (typically only 600s per polarizer)
there were few or no globular clusters that could be used for all frames,
especially as the chip orientation changed from epoch to epoch.  Extensive
testing gives us confidence, however, that our procedure successfully and
repeatably aligns the images to $\pm 0.2$ pixels.    
Following MULTIDRIZZLE, the orientation was set so that the y-axis corresponds
to north.  The
final result is a cosmic ray rejected and geometrically corrected image for each
polarizer at each epoch \citep{fru09}. 

For the F606W data, it was necessary to subtract galaxy emission 
before performing photometry and polarimetry.
We first created a composite image of the galaxy by
drizzle-combining all epochs together, to improve the S/N on the host galaxy.  
We then modeled the galaxy emission using the ELLIPSE and BMODEL tasks in IRAF.
After the model image was computed, it was split into 3
models to correspond to each polarizer, and subtracted, using the IRAF command
IMCALC, from the corresponding drizzled image for each polarizer.

\subsection{Polarization Images}

Next, the drizzled images from each polarizer were used to create images for
Stokes I, Q, and U, along with their errors.  For the ACS data, we followed 
the procedure in the ACS data handbook \citep{May10}. 
For the WFPC2 data, the Stokes images were computed by using the WFPC2
Polarization Calibrator tool
\footnote{$http://www.stsci.edu/hst/wfpc2/software/wfpc2\_pol\_calib.html$}.
This produces the coefficients needed to compute the Stokes images by using
Mueller matrices that describe the pick-off mirror, the polarizer filter, and
the various rotations between the optical elements and the reference frames.
This tool is accurate to $\approx$1-2\% \citep{mcm97}. Both of these procedures
yield Stokes  $U$, $Q$ and $I$ images that are
combined in a standard way to produce emission weighted
fractional polarization (defined as $P=(Q^2+U^2)^{1/2}/I$) and 
electric vector position angle (defined as EVPA = $1/2 \times \tan^{-1}(U/Q)$) 
images. 


After the Stokes images are found, we accounted for the well-known Rician bias 
in $P$ (Serkowski 1962)  using a Python code adapted from the STECF IRAF package
(Hook et al. 2000).  This code debiases the $P$ image following Wardle \&
Kronberg (1974), and calculates the error in polarization PA, accounting for the
non-Gaussian nature of its distribution (see Naghizadeh-Khouei \& Clarke 1993).
In performing this calculation, pixels with signal to noise $(S/N)< 0.1$ were
excluded outright, and since the debiasing is done with a ``most-probable
value'' estimator, pixels where the most-probable value of $P$ was negative, or
above the Stokes $I$ value ({\it i.e.} $P > 100\%$) were blanked. 
This code was first used in Perlman et al. (2006). 

\begin{deluxetable*}{crrrrrrrr}
\tablecaption{Photometry of M87 Components} 
\tablewidth{0pt}
\tablehead{ \colhead{Date (No.)} & \multicolumn{4}{c}{Core Fluxes ($\mu$Jy)} & \multicolumn{4}{c}{HST-1 Fluxes ($\mu$Jy)}  \\
	& \colhead {F606W} &\colhead{F330W}& \colhead{F250W} & \colhead{F220W} &\colhead {F606W} & \colhead{F330W} & \colhead{F250W} & \colhead{F220W} }

\startdata 
Dec 07 2002~ (1) &$   671 \pm 7  $& $305 \pm 2 $ &  ....           &$ 146 \pm 11 $ & $ 232 \pm 2 $ &$  231\tablenotemark{a} \pm 2$&       ...        &$    141\tablenotemark{b} \pm 6 $ \\ 		       
Dec 10 2002~ (2) &$   630 \pm 6  $&$ 305 \pm 2 $ &$      ....     $&$ 146 \pm 11 $ & $ 237 \pm 2 $ &$  231\pm 2 $ &$ ....   	    $&$    141 \pm 6   $\\	       
Nov 29 2003~ (3) &$   478 \pm 5  $&	 ...	 &  	  ....     &$ 137 \pm 11 $ & $ 427 \pm 4 $ &   .... 	&$      ....        $&$    244 \pm 7   $\\		   
Nov 28 2004~ (4) &$  1066 \pm 11 $&$  475 \pm 2$ &  $306 \pm 14   $&$ 226 \pm 14 $ & $ 970 \pm 10$ &$  882 \pm 4 $&$     674 \pm 12   $&$    631 \pm 12  $\\
Dec 26 2004~ (5) &$  1306 \pm 13 $&    ....      &$    363 \pm 10 $&  .... 	   & $1113 \pm 11$ &   .... 	&$     719 \pm 10   $&$       ....     $\\	       
Feb 09 2005~ (6) &$   891 \pm 9  $&    ....      &$    280 \pm 9  $&     .... 	   & $1224 \pm 12$ &   ....	&$   738 \pm 10     $&$       ....     $\\			       
Mar 27 2005~ (7) &$  1037 \pm 10 $&    ....      &$    328 \pm 10 $&     .... 	   & $1404 \pm 14$ &   ....	&$   903 \pm 11     $&$       ....     $\\		       
May 09 2005 ~(8) &$   932 \pm 9  $&$ 446 \pm 3  $&$274 \pm 9	  $&$  217 \pm 9 $ & $1333 \pm 13$ &$  1209 \pm 5$&$   878 \pm 14     $&$    853 \pm 14  $\\
Jun 22 2005 ~(9) &$   839 \pm 8  $&    ....      &$    273 \pm 9  $&     .... 	   & $1150 \pm 12$ &   .... 	&$   877 \pm 11     $&$       ....     $\\		       
Aug 01 2005~(10) &$   639 \pm 6  $&     ....     &$    192 \pm 7  $&     .... 	   & $1117 \pm 11$ &   ....	&$   664 \pm 10     $&$       ....     $\\		       
Nov 29 2005~(11) &$   735 \pm 7  $&$  349 \pm 2 $&$217 \pm 8	  $&$  170 \pm 8 $ & $1019 \pm 10$ &$   872 \pm 6$&$   604 \pm 12     $&$    616  \pm 12 $\\
Dec 26 2005~(12) &$   756 \pm 8  $&    ....      &$    234 \pm 8  $&     .... 	   & $ 987 \pm 10$ &   ....	&$   590 \pm 9      $&$       ....     $\\	       
Feb 08 2006~(13) &$   631 \pm 6  $&   ....       &$    201 \pm 8  $&$  160 \pm 8 $ & $ 771 \pm 8 $ &   ....	&$   482 \pm 8      $&$    469 \pm 8   $\\
Mar 30 2006~(14) &$   780 \pm 8  $&   ....       &$    232 \pm 8  $&	  ....     & $ 656 \pm 7 $ &   ....	&$   404 \pm 7      $&$       ....     $\\		       
May 23 2006~(15) &$   862 \pm 9  $&$  372 \pm 2 $&$ 193 \pm 7	  $&$ 156.4 \pm 7$ & $ 592 \pm 6 $ &$   488 \pm 4$&$   361 \pm 7      $&$    357 \pm 7   $\\
Nov 28 2006~(16) &$  1370 \pm 14 $&$  636 \pm 3$ &$323 \pm 9	  $&     .... 	   & $ 862 \pm 9 $ &$   720 \pm 4$&$   486 \pm 10     $&$        ....    $\\
Dec 30 2006~(17) &$  1006 \pm 10 $&    ....      &$    276 \pm 9  $&     .... 	   & $ 682 \pm 7 $ &$  ....      $&$   414 \pm 7      $&$        ....    $\\
Nov 25 2007~(18) &$  1292 \pm 17 $&    ....      &$       ....	  $&     .... 	   & $ 372 \pm 4 $ &$  ....      $&$    ....          $&$       ....     $\\
\enddata
\tablenotetext{a}{Observations taken Dec. 10, 2002.}
\tablenotetext{b}{Observations taken Nov. 30, 2002.}
\end{deluxetable*}

\begin{deluxetable*}{crrrrcrrrrc}
\tablecaption{Polarimetry and Spectral information}
\tablewidth{0pt}
\tablehead{\colhead{No.} & \multicolumn{5}{c}{Core} & \multicolumn{5}{c}{HST-1} \\
& \colhead{F606W}& \colhead{F606W}& \colhead{F330W} & \colhead{F33OW} && \colhead{F606W}& \colhead{F606W} & \colhead{F330W} & \colhead{F33OW} \\
& \colhead{$P$(\%)} & \colhead{EVPA($^\circ$)} & \colhead{$P$(\%)} & \colhead{EVPA($^\circ$)} & \colhead{$\alpha_{O-UV}$}&\colhead{$P$(\%)} & \colhead{EVPA($^\circ$)} &\colhead{$P$(\%)} &\colhead{EVPA($^\circ$)} & \colhead{$\alpha_{O-UV}$}
}
\startdata 
1  & $ 3.1 \pm  0.3 $ & $ -79.9 \pm 3.12 $ & $ 4.5 \pm 0.7 $ & $ -82.5 \pm 4.7 $ & $ 1.50 \pm 0.06$ & $40.2 \pm 4.0$ & $-65.8\pm3.0$ & $ 34.0\tablenotemark{a} \pm 3.5 $ &  $-67.5 \pm 3.1$  & $ 0.07 \pm 0.04 $ \\	   
2  & $ 3.7 \pm  0.4 $ & $ -79.6 \pm 3.17 $ & $ 4.5 \pm 0.7 $ & $ -82.5 \pm 4.7 $ & $ 1.41 \pm 0.04$ & $39.9 \pm 4.0$ & $-65.3\pm3.0$ & $ 34.0 \pm 3.5 $ &  $-67.5 \pm 3.1$ & $ 0.12 \pm 0.04 $ \\
3  & $ 5.5 \pm  0.6 $ & $ -70.3 \pm 3.07 $ & 	 ....	     &	   ....     & $ 1.23 \pm 0.08$ & $39.9 \pm 4.0$ & $-51.3\pm3.0$ &        ....      &	....	  & $ 0.58 \pm 0.03 $ \\    
4  & $ 1.4 \pm  0.2 $ & $ -10.4 \pm 3.37 $ & $13.4 \pm 1.5 $ & $  73.7 \pm 3.4 $ & $ 1.54 \pm 0.06$ & $23.4 \pm 2.3$ & $-58.2\pm3.0$ & $ 24.8 \pm 2.6 $ &  $-64.7 \pm 3.1$ & $ 0.32 \pm 0.03 $ \\
5  & $ 2.0 \pm  0.3 $ & $ -55.4 \pm 3.69 $ &	 ....	     &	   ....     & $ 1.58 \pm 0.04$ & $27.7 \pm 2.8$ & $-56.9\pm3.0$ &        ....      &	....	  & $ 0.57 \pm 0.02 $ \\   
6  & $ 2.2 \pm  0.3 $ & $ -98.9 \pm 3.76 $ &	 ....	     &	   ....     & $ 1.42 \pm 0.04$ & $36.7 \pm 3.7$ & $-58.4\pm3.0$ &        ....      &	....	  & $ 0.66 \pm 0.02 $ \\   
7  & $ 6.1 \pm  0.6 $ & $ -53.2 \pm 3.1  $ &	 ....	     &	   ....     & $ 1.41 \pm 0.04$ & $42.5 \pm 4.3$ & $-61.5\pm3.0$ &        ....      &	....      & $ 0.57 \pm 0.02 $ \\   
8  & $ 4.1 \pm  0.4 $ & $ -78.9 \pm 3.05 $ & $10.0 \pm 1.3 $ & $ -86.8 \pm 3.8 $ & $ 1.42 \pm 0.04$ & $34.5 \pm 3.5$ & $-64.7\pm3.0$ & $ 38.1 \pm 3.9 $ &  $-62.7 \pm 3.0$ & $ 0.37 \pm 0.02 $ \\
9  & $ 10.7\pm  1.1 $ & $ -36.0 \pm 3.04 $ &	 ....	     &	   ....     & $ 1.37 \pm 0.04$ & $32.4 \pm 3.2$ & $-63.8\pm3.0$ &        ....      &	....	  & $ 0.35 \pm 0.02 $ \\   
10 & $ 8.6 \pm  0.9 $ & $ -62.0 \pm 3.09 $ &	 ....	     &	   ....     & $ 1.48 \pm 0.05$ & $36.4 \pm 3.7$ & $-60.6\pm3.0$ &        ....      &	....	  & $ 0.68 \pm 0.02 $ \\   
11 & $ 6.6 \pm  0.7 $ & $ -69.1 \pm 3.03 $ & $10.1 \pm 1.4 $ & $ -77.2 \pm 4.0 $ & $ 1.45 \pm 0.05$ & $32.6 \pm 3.3$ & $-62.2\pm3.0$ & $ 29.3 \pm 3.0 $ &  $-62.0 \pm 3.0$ & $ 0.45 \pm 0.03 $ \\
12 & $ 4.4 \pm  0.5 $ & $ -90.0 \pm 3.26 $ &	 ....	     &	   ....     & $ 1.44 \pm 0.05$ & $28.1 \pm 2.8$ & $-62.4\pm3.0$ &        ....      &	....	  & $ 0.67 \pm 0.02 $ \\   
13 & $ 2.5 \pm  0.3 $ & $ -69.4 \pm 3.24 $ &	 ....	     &	   ....     & $ 1.39 \pm 0.07$ & $26.3 \pm 2.6$ & $-59.5\pm3.0$ &        ....      &	....	  & $ 0.55 \pm 0.03 $ \\
14 & $ 2.6 \pm  0.3 $ & $ -71.4 \pm 3.7  $ &	 ....	     &	   ....     & $ 1.50 \pm 0.05$ & $24.1 \pm 2.4$ & $-59.0\pm3.0$ &        ....      &	....	  & $ 0.63 \pm 0.03 $ \\    
15 & $ 1.0 \pm  0.1 $ & $ -1.2  \pm 3.91 $ & $ 3.0 \pm 1.0 $ & $ -25.7 \pm 9.8 $ & $ 1.63 \pm 0.05$ & $20.7 \pm 2.1$ & $-62.7\pm3.0$ & $ 24.4 \pm 2.7 $ &  $-50.2 \pm 3.3$ & $ 0.47 \pm 0.03 $ \\
16 & $ 6.6 \pm  0.7 $ & $ -39.1 \pm 3.01 $ & $ 2.0 \pm 0.7 $ & $ -57  \pm 10   $ & $ 1.50 \pm 0.04$ & $34.9 \pm 3.5$ & $-66.1\pm3.0$ & $ 34.8 \pm 3.6 $ &  $-64.2 \pm 3.1$ & $ 0.48 \pm 0.03 $ \\
17 & $ 10.2\pm  1.0 $ & $ -46.4 \pm 3.04 $ &	 ....	     &	   ....     & $ 1.60 \pm 0.04$ & $20.3 \pm 2.1$ & $-58.7\pm3.0$ &        ....      &	....	  & $ 0.73 \pm 0.03 $ \\
18 & $ 6.7 \pm  3.0 $ & $ -130.1\pm 1.30 $ &	 ....	     &	   ....     &        ....      & $22.2 \pm 3.0$ & $-76.5\pm0.7$ &        ....      &	....	  &	  .....       \\
\enddata
\enddata
\tablenotetext{a}{Observations taken Dec. 10, 2002.}
\end{deluxetable*}

\subsection{Aperture Photometry and Polarimetry}

To obtain fluxes in the Stokes parameters,  we performed aperture photometry.   
For the ACS data, we used two concentric apertures for the core 
(one with a radius of 11 pixels,
the other with a radius of five pixels) and one centered on HST-1 with a radius
of 11 pixels.  The smaller of the two core apertures excludes a knot that
emerged from the core in later images, while the larger core aperture includes
it.  While the  two lightcurves show the same behavior, we use only the smaller
aperture in  this paper.  For the lone WFPC2 dataset (epoch 18) we used
apertures of 4 pixels radius for both the core and HST-1, and also excluded in 
each aperture an annular region between 4-6 pixels from the other.  This
alternate procedure was made necessary by the much larger pixels of the
WFPC2/WFC chips.  
For the F606W images, after galaxy subtraction
rectangular regions located quasi-randomly throughout the galaxy were used to 
verify the flatness of the remaining background.  A similar strategy was used 
for the F330W images, where the galaxy contributon was minimal. 
Aperture correction was done to account for the wide wings
of the HST point spread function (PSF).  As the nucleus and HST-1 are either
unresolved or nearly so,  we modeled them as PSFs, using  TINYTIM
\citep{May10,boh07}.  The aperture corrections were generally 10-20\% in flux,
similar to those found by Sirianni et al. (2005); however, our method allowed us
to account for the fact that the position of these components varied widely on
the chip from observation to observation.  Count rates were converted to  flux
using SYNPHOT, recomputing PHOTFLAM values by assuming a spectral index
$\alpha=0.7$ ($F_\nu \propto \nu^{-\alpha}$), appropriate for most of the M87
jet (Perlman et al. 2001).  Typical errors for this procedure are $<5\%$.
Finally, we also accounted for Galactic extinction using data from NED, which
gives $A_B=0.096$,    as well as standard extinction curves.  The resulting
fluxes in Stokes I are listed in Table 1, while the fractional polarizations and
EVPA are given in Table 2.

High-quality UV photometry of these two components was published
recently by Madrid (2009).   We have utilized those measurements, albeit with
some modifications.  Rather than correcting all fluxes to a single wavelength
as in the Madrid paper, we have utilized the uncorrected fluxes in both F220W
and F250W.  This minimizes the number of assumptions, and  
also allows us to make use of all the data in the 6 epochs
where observations were taken in both bands, increasing the accuracy of the
measured spectral indices in those epochs and also allowing us to check for 
significant emission in the Mg II $\lambda$ 2798 line, which falls near the 
center of the F250W bandpass, but is outside the F220W bandpass (no evidence
of this line emission was found).  The reader
is referred to that paper for reduction steps.  We list the
resulting fluxes in Table 1.  
In Table 2 we list the spectral indices $\alpha_{UV-O}$.

Errors were propagated in both datasets using standard techniques. The
propagated errors include Poisson errors, an additive noise term (the RMS
background calculated post galaxy subtraction) and the read-out and
discretization noises, as well as an additional 1\% term to account for
uncertainties in SYNPHOT models (Bohlin et al. 2007b).  For this analysis, we
have ignored errors in the F606W 
galaxy model. The resulting uncertainties in the
Stokes $Q$ and $U$ images are approximately Gaussian in nature, with their
values being approximately equal to the sum in quadrature of the individual
polarizer image errors. Hence, Gaussian error propagation for $P$ is appropriate
for our purposes.  


\section{Results}

Figures 1 and 2 show our results for the total flux, optical-UV spectral index,
fractional polarization and EVPA variations of both the nucleus and HST-1.     
The total fluxes shown are in the F606W band, while all other panels use all 
available data.  The total  flux variations mirror those previously shown
for the near-UV by Madrid (2009),  showing the very large flare in HST-1 as well
as two smaller flares in the nucleus during the same time.   Because the large
majority of these observations occurred during 2004-2006, when  the knot was
already very bright, we do not see the full dynamic range of the variability
exhibited by knot HST-1 during its flare -- nearly a factor 150 at 2500 \AA,
where its flux increased from 6 $\mu$Jy in 1999 to a peak of 854 $\mu$Jy in
early 2005. We used our data to measure the distance between the two features,
which in all epochs is consistent with $0.885 \pm 0.010$ arcsec, with no
evidence of motion.  This sets an upper limit of 1.56 pc ($2 \sigma$)  on
positional change of the flux maximum of HST-1 during the  roughly 5-year
timespan covered by these data.   

\begin{figure}
\includegraphics[scale=0.44]{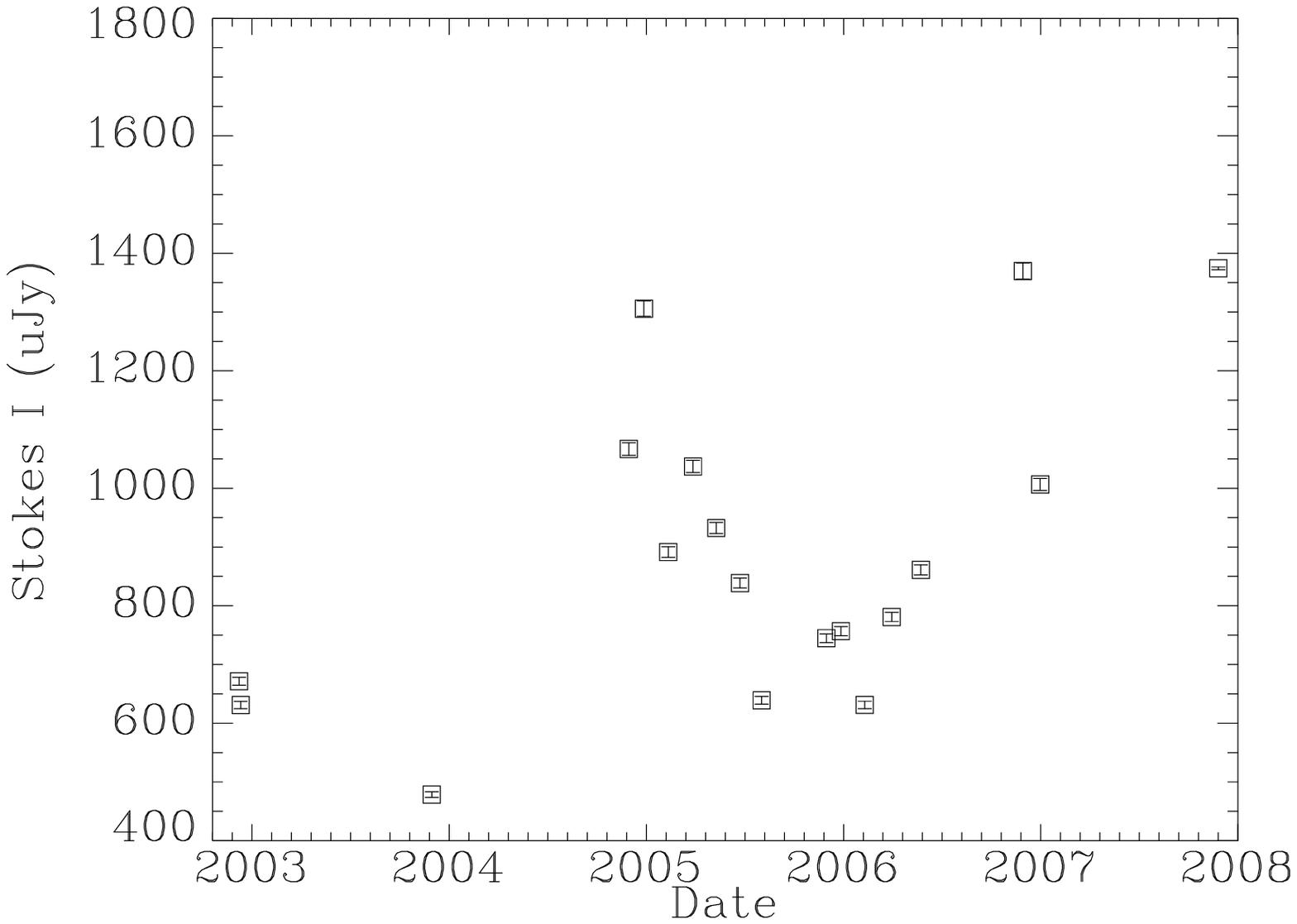}

\includegraphics[scale=0.44]{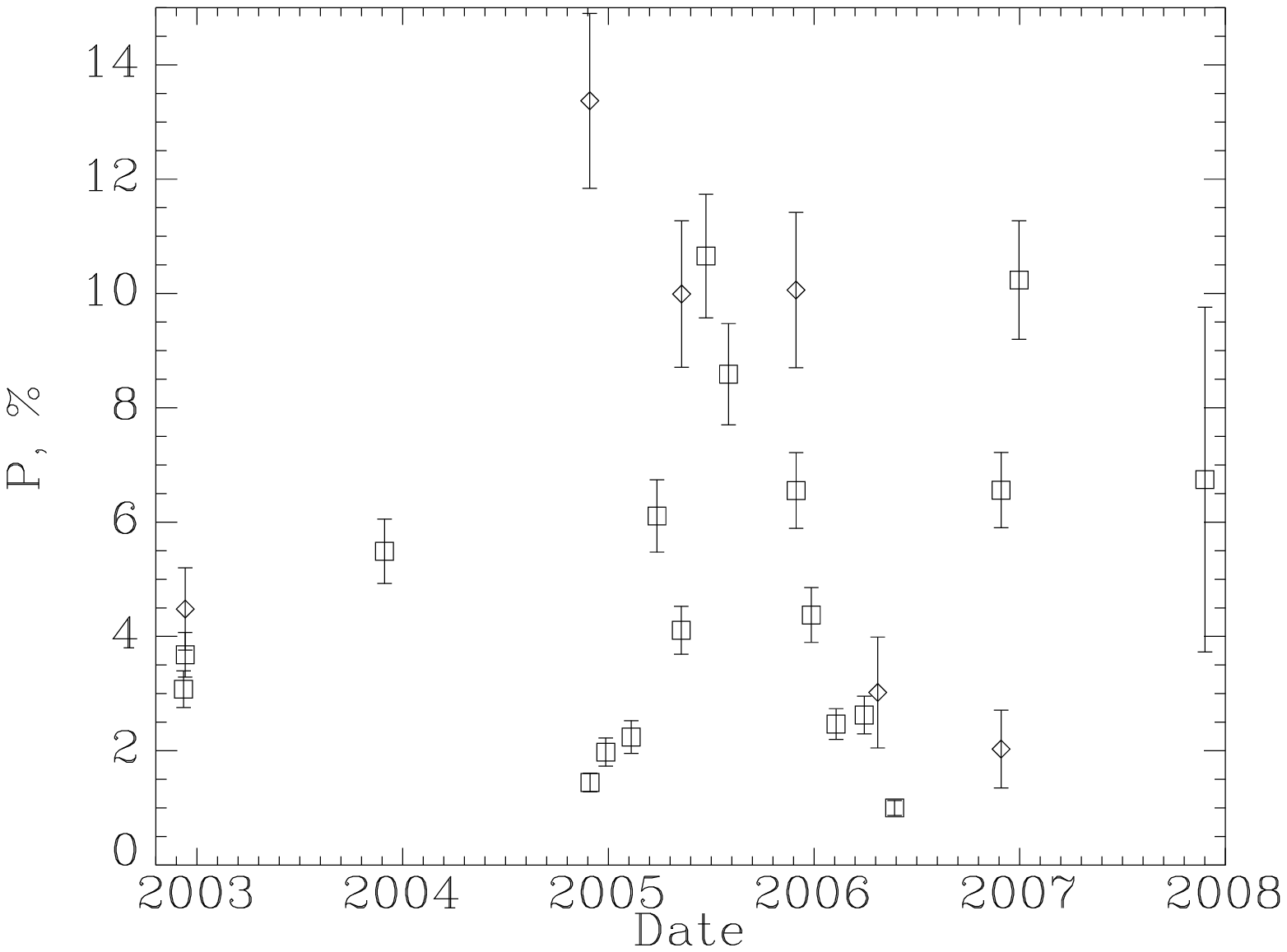}

\includegraphics[scale=0.44]{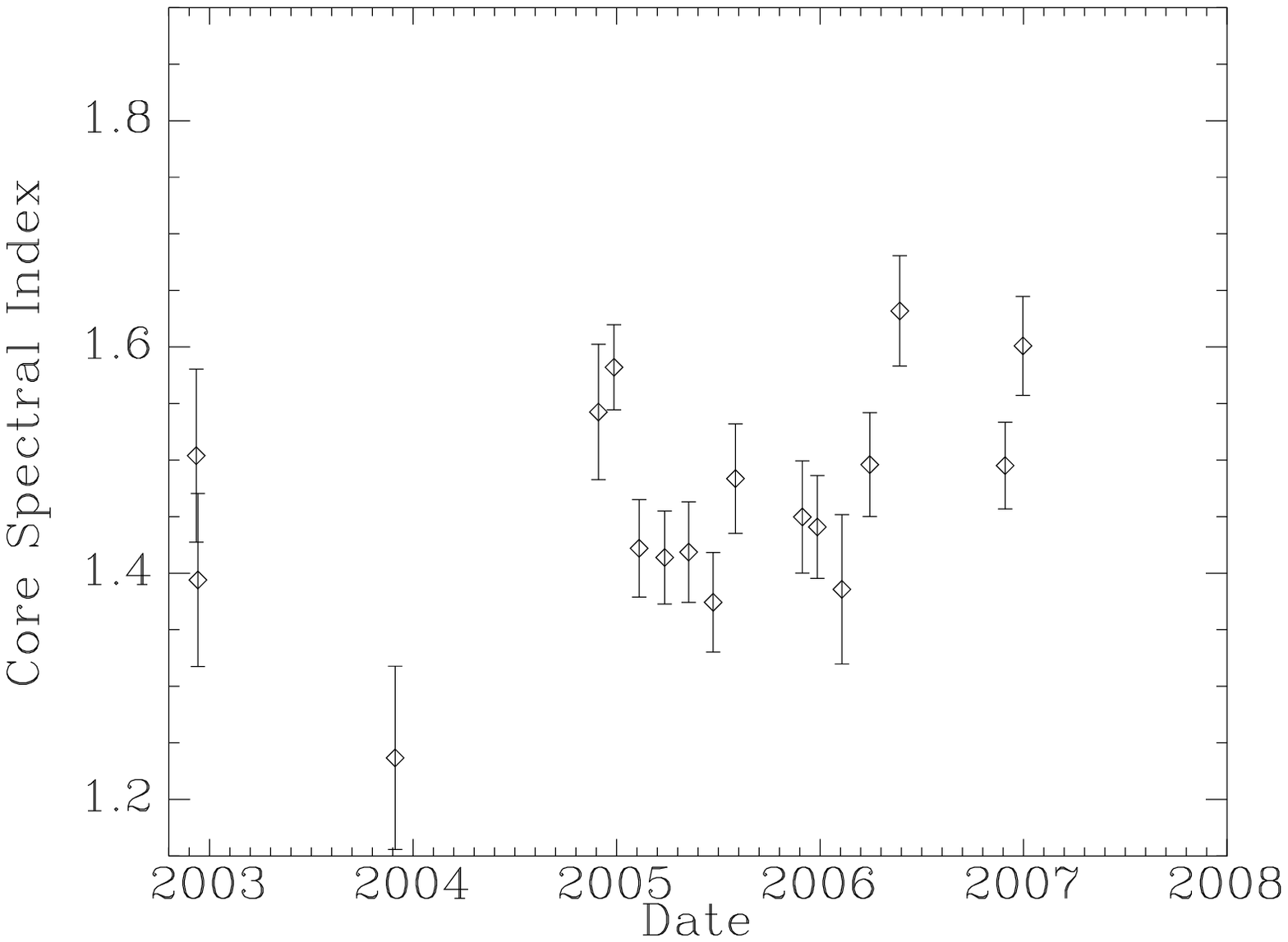}

\includegraphics[scale=0.44]{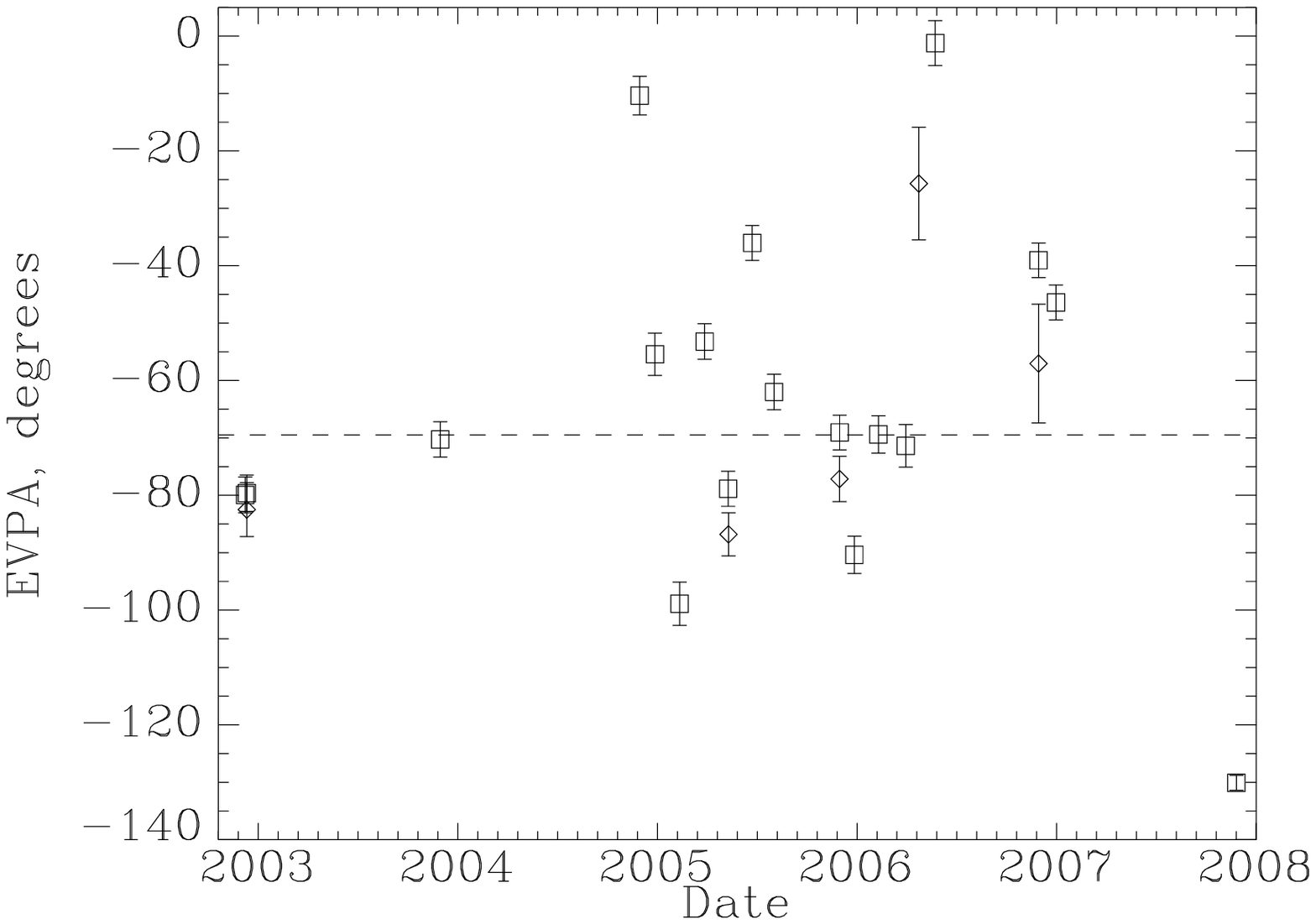}

\caption{Variations in Total flux (i.e., Stokes I), 
fractional polarization, spectral index and EVPA, are plotted
for the nucleus of M87.  A dashed line in the EVPA panel reflects the jet PA.
The F606W observations are plotted as squares, while in the second and fourth
panel the F330W polarimetry is plotted as diamonds.
See \S 3.2 and 4 for discussion.}

\label{Flux Core}
\end{figure}

\begin{figure}
\includegraphics[scale=0.44]{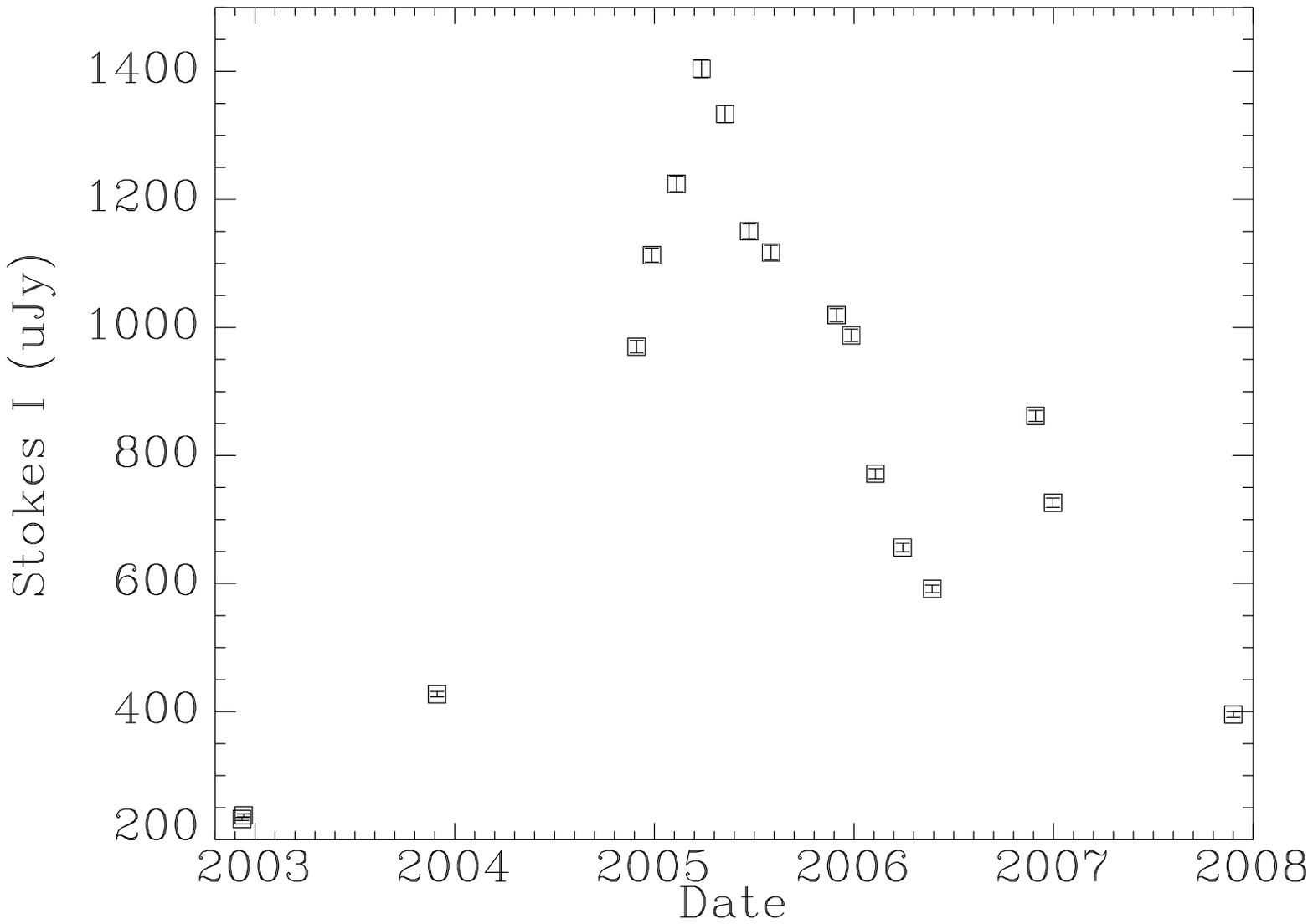}
\includegraphics[scale=0.44]{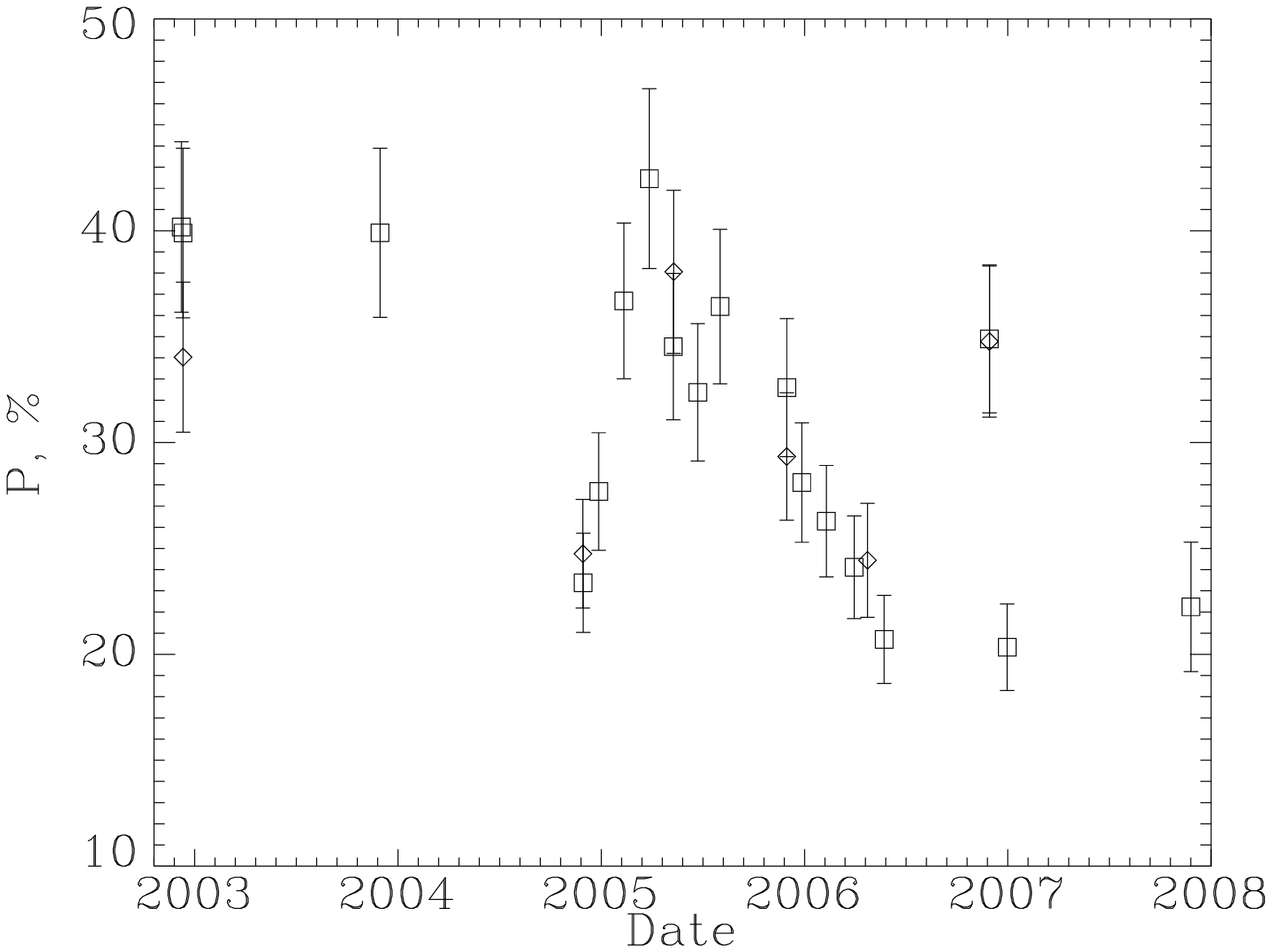}

\includegraphics[scale=0.44]{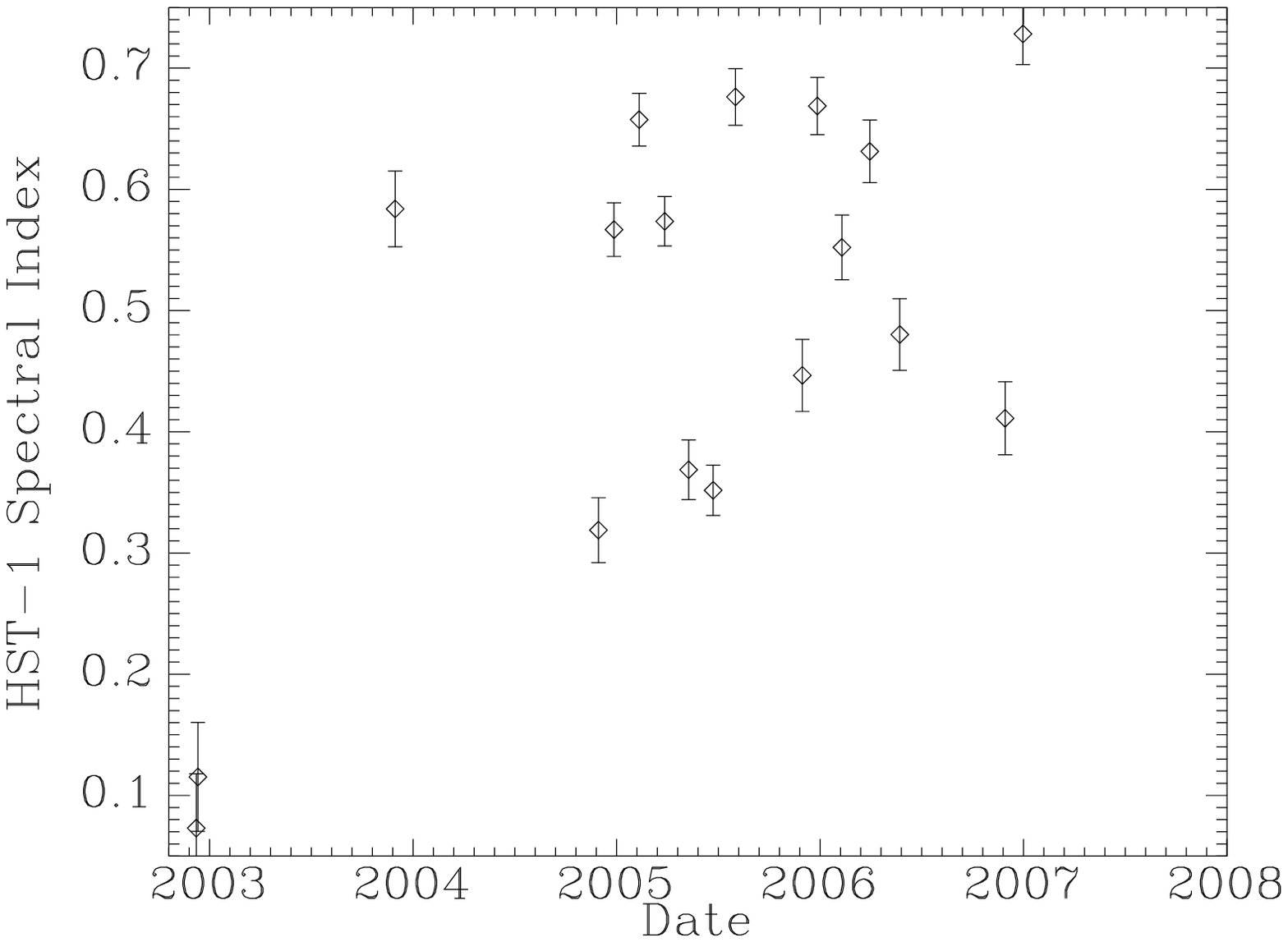}
\includegraphics[scale=0.44]{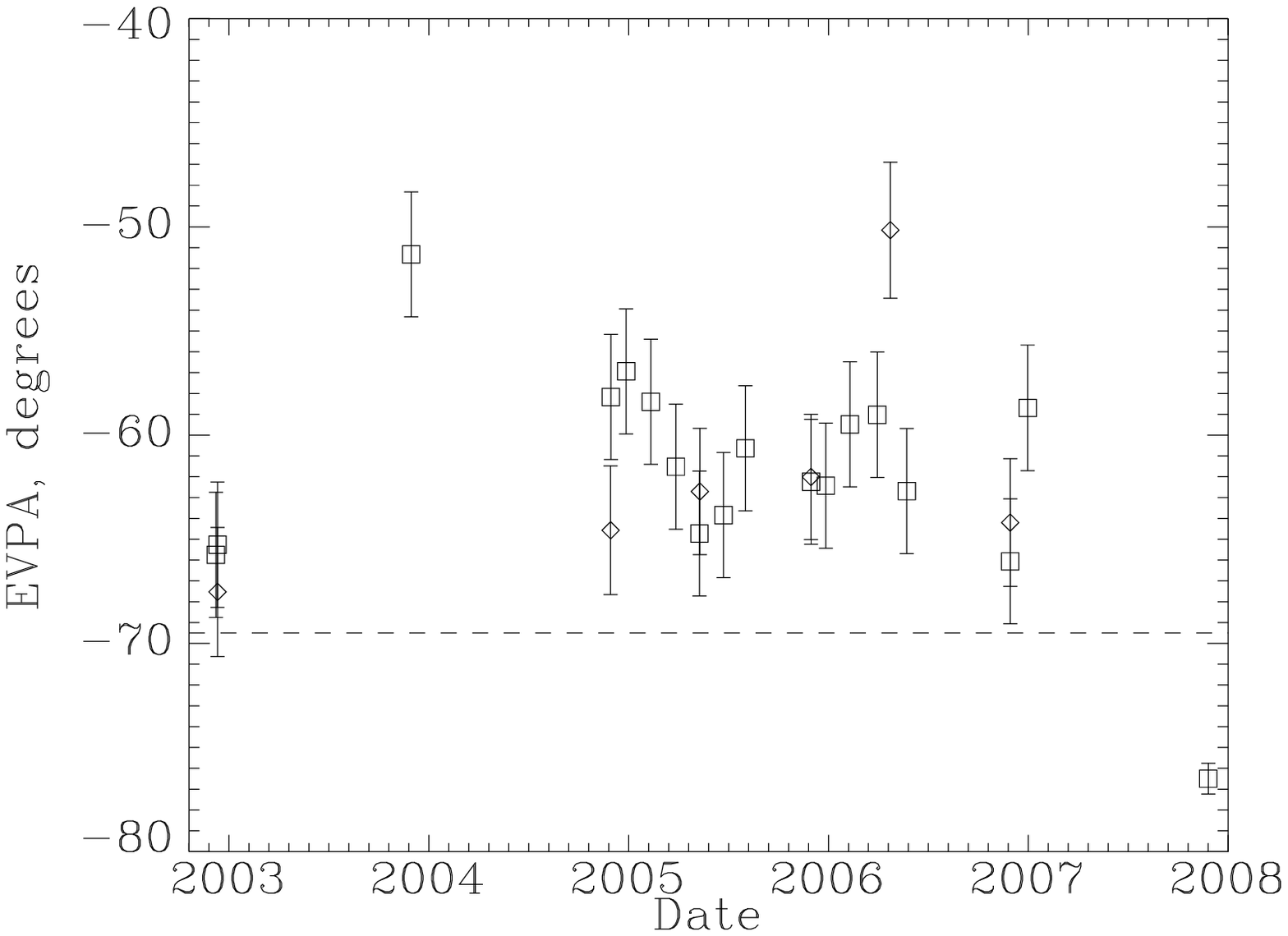}

\caption{Variations in Total flux, 
fractional polarization, spectral index and EVPA, are plotted
for knot HST-1.  A dashed line in the EVPA panel reflects the jet PA.  
The F606W observations are plotted as squares, while in the second and fourth
panel the F330W polarimetry is plotted as diamonds.
See \S 3.2 and 4 for discussion.}

\label{Flux HST 1}
\end{figure}

Comparing Figures 1 and 2 we see that the nucleus and HST-1 display very
different characteristics in polarization and spectral index variability.
Both display large changes
in polarization characteristics during the six-year timespan of our
observations. The nucleus is seen to range between 1-13\% polarization during
this time, with the EVPA changing by as much as 90 degrees. 
HST-1 is much more highly polarized, with its 
polarization ranging from 20-45\%, but much less variability 
(marginally significant at most) in the EVPA, which in our data ranges from 
roughly $-50^\circ$ to $-75^\circ$, with a typical value of 
$\sim -62 ^\circ$, 
about 7-8 degrees away from the nominal jet direction of $-69.5^\circ$ but 
rather closer (4 degrees) to the average PA of the radius vector
for HST-1 during VLBA
observations (Paper IV) of $-66^\circ$.  The difference between the two angles
is only 1 $\sigma$ for any one point,  but in 17 of 18 epochs the EVPA is
significantly displaced towards the north from the radius vector, making the
difference significant at roughly  the $3 \sigma$ level (see also \S 5.1).  

For the most part the polarization characteristics in F330W track the ones seen
in F606W.  This is true for all epochs for knot HST-1.  Thus that knot displays
no evidence for frequency-dependent polarization behavior.  For the nucleus,
however, one epoch shows significant ($>2 \sigma$)  differences between
polarization characteristics in the two bands, namely 28 Nov 2004.  As can  be
seen in Figure 1, this difference is highly significant both in $P$ (8 $\sigma$)
as  well as EVPA (10 $\sigma$), and we have eliminated all possible sources of
instrumental  error in the F330W and F606W data for this epoch.  We note that
this epoch is very near the peak of a flare in the nucleus; however, this flare
does not exhibit a spectrum that is significantly different than surrounding
epochs.  Unfortunately the next epoch does not have UV polarimetry so we are
unable to  comment further on whether any frequency-dependent polarization
pattern developed during this flare.  

Another perspective can
be gained by looking at relationship between flux and
fractional polarization.   This is done in
Figure 3, using the F606W data only.  
In knot HST-1 (bottom panel), this reveals a strong correlation 
between total flux and fractional polarization, particularly between 2004 
November and 2006 December (epochs 4-17), when the flux variability was 
dominated by the main part of the flare and the monitoring was most frequent.  
The other four points represent times where the flux variability was either 
dominated by or had a significant contribution from smaller scale events. 
Using only epochs 4-17, 
a Spearman's rho-test indicates $\rho=0.842$ and 
$P=1.6 \times 10^{-4}$.  
Inspection of the F330W data shows they also follow this correlation.
During this time the EVPA is nearly 
constant (Figure 1), although there is possible evidence for a quasi-sinusoidal
modulation. This gives the clear impression of the variability being 
dominated by a single component with a highly ordered magnetic field.  We
discuss the implications further in \S 5.2.
   

A pattern is much more difficult to pick out for the nucleus.  We have therefore
added red arrows to Figure 3 to guide the eye  during the part of the campaign
where there is evidence of a pattern.  During epochs 6-15,  there was a
coherent increase and then decrease in the fractional polarization of the
nucleus, at a time when the flux variations where small (Figure 1).   This
translates into a  near-vertical ``loop'' in the polarization-flux plane (Figure
3, {\it top panel}), with the initial increase in polarization being correlated
with the decrease in flux from epoch 5's maximum, and then a nearly monotonic
decrease being observed over epochs 9-15 while the flux varied by only 25\%. 
This is discussed further in \S 5.3.

\begin{figure}[h]
\includegraphics[scale=0.42]{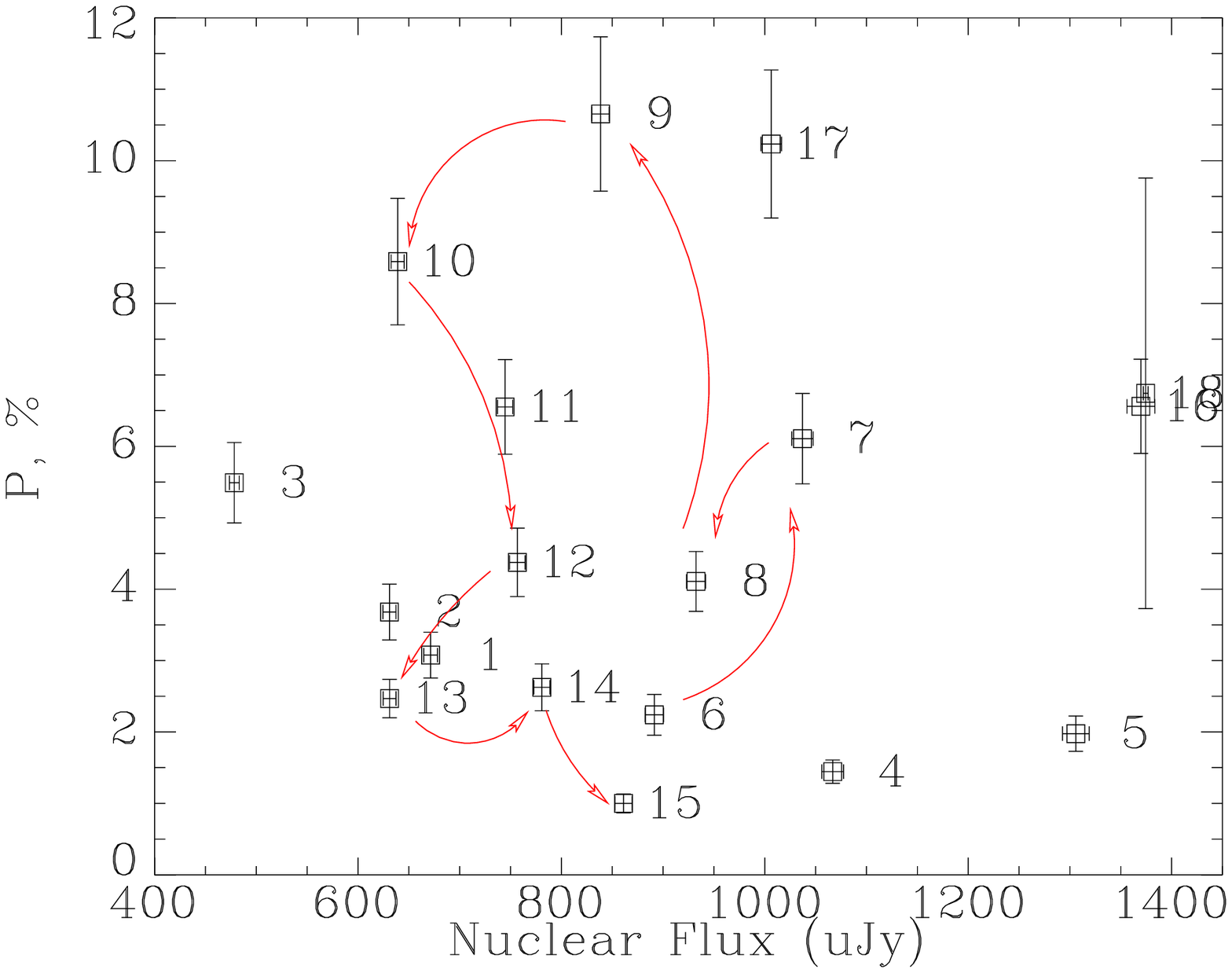}
\includegraphics[scale=0.5]{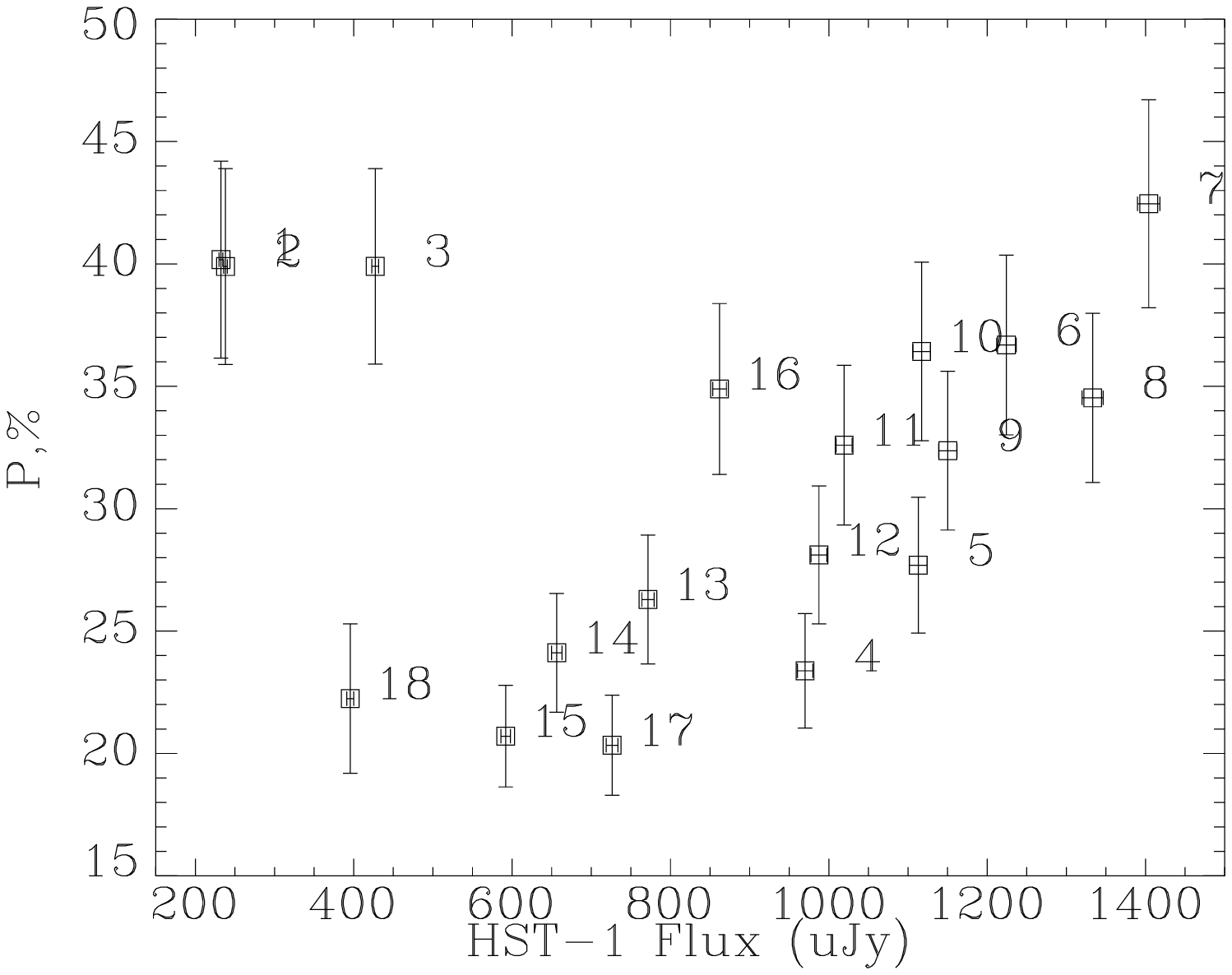}
\label{Fractional Polarization vs. Flux Core}

\caption{Graphs of fractional polarization versus flux for the core (top) and 
HST-1 (bottom). The observation epochs have been labelled sequentially.  Notice
that for HST-1, epochs 4-18 display a very strong correlation between the 
flux and polarization, whereas the behavior for the nucleus is very different,
with a 'loop' seen between epochs 9-15, but otherwise no organized pattern.}

\label{Fractional Polarization vs. Flux HST 1}
\end{figure}

Table 2 and Figures 1-2 also describe the evolution of the UV-optical spectral 
index $\alpha_{UV-O}$.   As can be seen, $\alpha_{UV-O}$ behaves differently for
the nucleus than it does for HST-1.  For the core, we do not see significant 
variability in the optical spectral index. By contrast, the spectral index is
strongly variable for HST-1.  For most of the  time, the variability of the
spectral index appears uncorrelated with flux, but during the brightest part of
the flare   (epochs 4-9, denoted on Figure 4 by red arrows), we see that the
spectral index is larger (i.e., steeper) when the flux is higher.  As can be
seen in Figure 4, during this time HST-1 describes a definite 'loop' in the
flux-spectral index plane.   This pattern has been called ``hard lags'' in the
blazar and optical variability literature (see e.g., Zhang 2002, Zhang et al.
2002; Fiorucci, Ciprini \& Tosti 2004; the term ``counterclockwise looping'' is
also in use, but note that those papers use the opposite sign convention for
$\alpha$ than we do). Looping in the other direction is seen for epochs 13-17
(denoted on Figure 4 by green arrows), e.g., ''soft lagging'' (Zhang et al.
2002), with oscillations in the plane in between (denoted on Figure 4 by blue
arrows).  Both ``looping'' patterns are known to arise
for particular relations between the  acceleration and  cooling timescales
controlling the spectral evolution of the radiating particles (in the framework
of simplified models for such an evolution) and are discussed  further in \S
5.2.

\begin{figure}[h]
\includegraphics[scale=0.49]{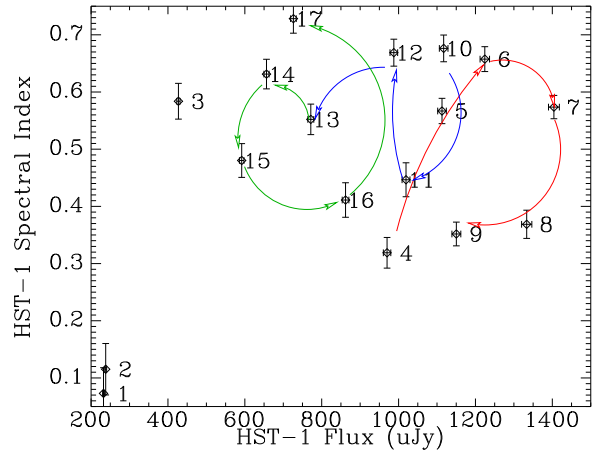}
\label{spectralvar}

\caption{The evolution of $\alpha_{UV-O}$ versus flux for knot HST-1.
Each epoch has been labelled sequentially, as in Figure 3.  Notice
the 
strong 'looping' behavior during the maximum of the flare in HST-1, as 
well as after (green and red arrows).  See \S\S 4, 5.1
for discussion.}

\end{figure}

\section{Discussion}


It is interesting to explore further the reasons behind the highly disparate
behaviors of the nucleus and HST-1, as seen in our data as well as other bands.
X-ray variability of the nucleus and HST-1 was studied in Papers I, III and V.  
Variability has also been seen at gamma-ray energies, where M87 has  
been detected in both TeV (Aharonian et al. 2003) 
and GeV (Abdo et al. 2009) energies (although, {\it n.b.}, in the gamma-rays
the M87 nucleus and jet cannot be resolved from one another).   
During the peak of the flare of HST-1 in March-May 2005, 
TeV variability on timescales of a few days (Aharonian et al. 2006) was seen, leading 
Stawarz et al. (2006) to associate the enhanced TeV emission seen in 2005 with the 
flare of HST-1.  However, the origin of other features in the TeV 
lightcurve is unclear. A major work
compiling all multi-wavelength and TeV gamma-ray variability data for M87 was
recently completed by Abramowski et al. (2011).   That work
concluded that while it remains plausible that both the unresolved
nucleus and HST-1 contribute to the TeV emission observed from M87 system during
the quiescence epochs and also the 2005 flare, during the 2008 and 2010 epochs
of the enhaced $\gamma$-ray activity of the source the nucleus is more likely to
have contributed the majority of the TeV flux. 

In the radio, both Paper IV as well as Chen et al. (2011) studied the variability
of knot HST-1, with Paper IV using
VLBA data, while Chen et al. (2011) used data from the VLA, both from roughly 
2003-2007.  The latter work has
angular resolution comparable to HST and includes an analysis of polarization
data, and finds a variable $P$, EVPA and rotation measure, as well as a 
radio spectrum that softened  when the knot was brighter.  It is
difficult to compare their data to ours in detail because Chen et al. (2011)
used a subset of the available data, including 
only 4 datasets between 2004-2006 (and none in 2005), when HST-1 was most
active.  However their findings, while consistent with the idea of non-cospatial
radio and optically emitting particle populations (Perlman et al.
1999), are difficult to reconcile with the much longer radiative lifetimes of
radio synchrotron emitting particles, as well as the similar 
radio and  UV lightcurves.      

To further explore the physical implications of our results, it is necessary to
discuss the physics of shocks and other disturbances in jets, which may explain
the behavior we see.  As will be seen, we do this because we believe that both 
behaviors may be explained by such disturbances.   Following this, we will 
then present physical interpretations of the behaviors of knot HST-1 and 
the nucleus.

\subsection{Shocks, Helical Distortions and Polarization Variability}

The behaviors we see in both the nucleus and HST-1 are direct reflections of 
the physics in the emitting region.  Since the data we have analyzed in this 
paper includes fluxes, optical-UV spectra and also polarimetry, we have
information both on the interplay between particle acceleration and cooling,
as well as the magnetic field structure that was either associated with this
behavior or produced it.  In order to motivate the discussion herein, it is
useful to summarize the commonalities in the behavior of both regions and then 
discuss why those commonalities argue for an origin in shocks and waves.  Both
HST-1 and the nucleus exhibit coherent patterns in the $(I,P)$ plane.  In the
case of HST-1 the pattern is simple:  polarization is correlated strongly with
intensity, while at the same time the EVPA remains essentially constant very 
close to the PA of the jet.  By
comparison, in the nucleus we see a 'loop' in the $(I,P)$ plane, with somewhat 
more complicated EVPA behavior, featuring wild swings of up to $100^\circ$, 
albeit around a dominant orientation that is once again near the PA of the 
jet's radial motion vector.  
The fact that both HST-1 and the nucleus display coherent variability patterns in
(I,P) while maintaining a single, dominant orientation of EVPA indicates that
in both regions, the details of the local magnetic field structure are tightly 
related to the efficiency of the particle acceleration. That component must have
a reasonably well ordered magnetic field structure, particularly in the case of
HST-1 because of its very high polarization (see \S 5.2 for further elaboration
on this issue).   These behaviors are all consistent with having been produced
through shocks and/or waves, although the details of the physics may be different
in the two regions.  

What types of shocks may be consistent with these behaviors?  Perhaps the
simplest type of shock to discuss is localized, planar, and oriented 
along or near the jet perpendicular. This type of model,
often known as a ``Laing sheet'' due to the fact that shocks of this type
characteristically compress
an initially random magnetic field into a thin sheet with magnetic field along
the sheet, has been  investigated extensively in the literature, particularly in
Laing (1980), Hughes, Aller \& Aller (1985), and Kollgaard et al. (1990).   In
this model, the properties of a relativistic shock can be completely 
determined by a few factors.  Primary among these is its 
compression ratio, $k=\Gamma_d\beta_d/(\Gamma_u\beta_u)$
\citep{Laing:1980,Wardle:1994}, where $\Gamma_{u,d}$ is the bulk Lorentz factor
and $\beta_{u,d}$ is the bulk speed with the subscripts referring to upstream 
and downstream quantities respectively.   One can also think of the
compression ratio in terms of the compression of a unit length.
Ignoring the plasma
magnetization, the shock jump conditions relate the upstream and downstream 
speeds via $\beta_u\beta_d=1/3$ \citep{Landau:1987} for a plasma with a fully
relativistic equation of state.  The other factors that
describe the shock are the spectral
index $\alpha$, Doppler factor $\delta$ and viewing angle $\theta_{ob}$.  More
specifically, in  this model, Kollgaard et al. (1990) found that the degree of
polarization is given by (their equation (2); note that our convention for the
spectral index $\alpha$ is the opposite of the one in Kollgaard et al.)
\begin{equation}
P = {\dfrac{3+3\alpha}{5+3\alpha}} ~
{\dfrac{\delta^2(1-k^2)\sin^2\theta_{ob}}{2-\delta^2(1-k^2)\sin^2\theta_{ob}}}.   
\end{equation}
This model is obviously dependent on the geometry chosen for the jet magnetic
field and for the shock -- in particular ignoring any helical component to the 
field and a strong, perpendicular shock. Such a model can easily produce a 
correlation between flux and polarization, 
along with a roughly constant EVPA parallel to the jet direction, as seen in knot
HST-1.  We will discuss in \S 5.2 the application of this model to HST-1.   

If, however, a different type of disturbance is envisioned,
quite different polarization behavior can be produced.   
The first possibility we will consider is a time varying upstream speed, $\beta_u$.  
This is supported by the variable
superluminal speeds observed in AGN jets\citep[e.g.][]{Lister:2009}.  In this
second model, the varying $\beta_u$ is related to the downstream flow speed via
$\beta_u\beta_d=1/3$ which induces variability in the shock compression factor,
$k$.   To induce a variation in the shock compression factor, the stochastic variation
in $\beta_u$ is modeled as being sinusoidal:
\begin{align}
\beta_u(t)=\beta_{u,0}+A_{\beta}\sin{\left(\omega_{\beta}t-\phi_{\beta}\right)}
\label{beta}
\end{align}
The parameters $(\beta_{u,0}, A_{\beta},\omega_{\beta},\phi_{\beta})$ are free
parameters of the model, and $t$ is time.  If the jet contains a disordered field
component $B_r$, no ordered toroidal field, and an ordered poloidal field $B_p$,
then the polarization depends on the magnetic field through the ratio
$\xi=\left|B_p/B_r\right|$.  Since the process of transforming an initially
ordered large-scale field to a tangled one is 
probably governed by the kink instability 
\citep{Spruit:1997,Begelman:1998,Marscher:2009}, it is
natural to assume that the value of $\xi$ at the shock fluctuates in time. 
Similar to $\beta_u$, we model the $\xi$ variation as
\begin{align}
\xi(t)=\xi_{0}+A_{\xi}\sin{\left(\omega_{\xi}t-\phi_{\xi}\right)}.
\label{xi}
\end{align}
The parameters $(\xi_0, A_{\xi},\omega_{\xi},\phi_{\xi})$ are free parameters of
the model, and $t$ is time.  Equations (\ref{beta}) and (\ref{xi}) allow the
calculation of the post shock intensity, $I$, and polarization $P$
\citep[e.g.][]{Kollgaard:1990}:
\begin{align}
I&\propto\delta_d^{2+\alpha}k^{-2}B_r^2\left\{2+\left[3k^2\xi^2-(1-k^2)\right]\sin^2{\theta_{ob}'}\right\} \label{Ik} \\
P& \approx \frac{3+3\alpha}{5+3\alpha}\frac{\delta^2\left[(1-k^2)-3k^2\xi^2\right]\sin^2{\theta_{ob}}}{2-\delta^2(1-k^2)\sin^2{\theta_{ob}}+3\delta^2\xi^2k^2\sin^2{\theta_{ob}}}. \label{Pk}
\end{align}
Note that equation (\ref{Ik}) is an approximation as it results from integrating over 
the line of sight with $\alpha$ set to unity for convenience (Wardle et al. 1994), 
and that equation (\ref{Pk}) as expected reduces to equation (1) for
$\xi=0$.  The downstream Doppler factor,
$\delta=(\Gamma_d-\Gamma_d\beta_d\cos{\theta_{ob}})^{-1}$, raised to the power
$2+\alpha$ consistent with a steady jet \citep{Lind:1985}, is included in the
expression for intensity since the flow speed varies in the emission region which
is downstream of the shock.  In our convention, the EVPA is parallel to the jet
axis for $P>0$, and is perpendicular to it for $P<0$.  The factor $B_r^2$ in
equation (\ref{Ik}) is held as constant in our model. Otherwise, if used as
another time varying parameter, it would merely modify the amplitude of the
intensity fluctuations.

Another plausible configuration is one where 
the jet contains a significant helical component to its
magnetic field.  Such a scenario might be particularly operative in the nucleus,
since AGN jets are generally thought to be launched and collimated within
the central $\sim$ parsec by magnetic fields that dominate other sources of
pressure. Its applicability to regions further from the nucleus is less certain
because it is not known how far downstream from the launching site the
jet retains a magnetically dominant ordered helical field since jets are thought
to be unstable to the $m=1$ kink mode
\citep{Narayan:2009,Marscher:2009,Spruit:2010}.  Despite this theoretical
uncertainty, some observational evidence suggests that jets contain helical
magnetic fields on parsec scales: parsec-scale bulk acceleration
\citep{Vlahakis:2004}, Faraday rotation gradients
\citep{Asada:2002,Gabuzda:2004,Zavala:2005,Kharb:2009}, and observed asymmetries
in the transverse profiles of polarization, brightness and spectral index  
\citep{Clausen-Brown:2011}.  

\begin{figure}
	\centering
		{\includegraphics[width=2.3in]{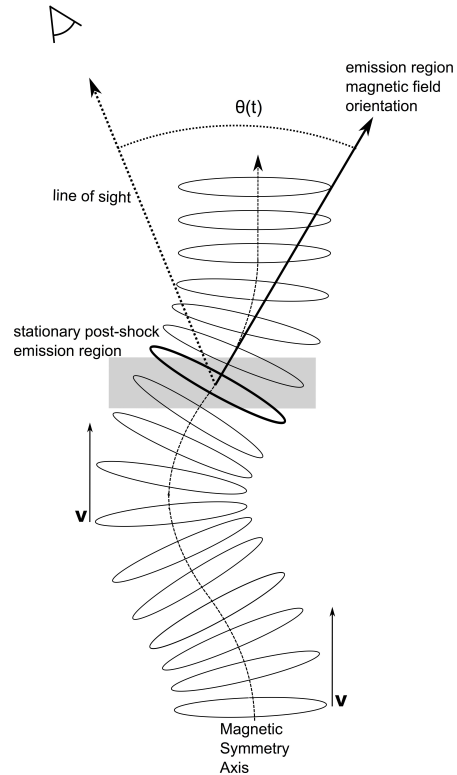}}
		\caption{Schematic of the helical distortion model illustrated 
		with toroidal loops of magnetic field.  As shown by the velocity 
		vectors (labeled by \textbf{v}), the velocity field is uniformly 
		directed despite the kinked jet.  As the kinked region of the 
		jet propagates through the standing shock's emission region 
		(gray region), the orientation of the magnetic field with 
		respect to the line of sight changes with time.  This change in 
		orientation is parametrized as $\theta'(t)$ in equation 
		(\ref{theta}).  The effects of shock compression, the poloidal 
		magnetic field and the change in the magnetic pitch angle, 
		$\psi'(t)$, are not shown.}
		\label{helical_cartoon}
\end{figure}

Let us assume that a component in the jet contains  a large-scale helical magnetic
field that is variable due to the growth of the $m=1$ kink mode.  
The jet's magnetic symmetry axis (parallel to the $B_z$ direction) will
then deform into a large scale helix that is carried with the jet's 
velocity field \citep{Mizuno:2011}, which we assume to be uniform in this work. 
If the helical deformation passes through a
standing shock, as shown in Figure \ref{helical_cartoon}, 
then the magnetic structure of the post-shock region will change
in time, thereby producing fluctuations in the post-shock synchrotron emission. 
Let us further assume that the ratio of $B_{\phi}'/B_z'$ varies in time in the
post-shock flow.  Kink mode simulations have found that, to avoid total jet
disruption, $B_{\phi}'/B_z'$ dynamically relaxes to $\sim 1$
\citep{Nakamura:2007}, while, in competition with this process, jet conical
expansion always increases the ratio.  Therefore, this competition will produce
fluctuations in $B_{\phi}'/B_z'$ at the standing shock.  

Within this scenario multiple types of disturbances can be envisioned.  Herein
we consider two.  Firstly, we may introduce sinusoidal variations both in the 
jet frame magnetic pitch angle $\psi'=\tan^{-1}(B_{\phi}'/B_z')$, and in the 
angle between the magnetic symmetry axis and the jet frame line of sight, 
$\theta'$, as shown in Figure \ref{helical_cartoon} :
\begin{align}
\psi'(t)&=A_{\psi}'\sin{\left(\omega_{\psi}t-\phi_{\psi}\right)}+\psi'_0 \label{psi} \\
\theta'(t)&=A_{\theta}'\sin{\left(\omega_{\theta}t-\phi_{\theta}\right)}+\theta_{ob}' \label{theta}
\end{align}
where $t$ is time and
$(A_{\theta}',A_{\psi}',\omega_{\theta},\omega_{\psi},\phi_{\theta},\phi_{\psi},
\theta_{ob}', \psi_0')$ are parameters of the model.  The jet frame viewing
angle, $\theta_{ob}'$, is actually set by the relation to the observer frame
viewing angle by $\sin{\theta_{ob}'}=\delta\sin{\theta_{ob}}$.
However, as the
Doppler factor for the inner jet is unconstrained, $\theta_{ob}'$ is treated as a
free parameter as long as the required Doppler factor is within reasonable
bounds.  It should be noted that when the magnetic field passes through the
standing shock, the field components lying in the shock plane will be amplified
by shock compression.  However, this only modifies the form of $\psi'(t)$ and
$\theta'(t)$; it does not prevent quasi-periodic variations from occurring. 

To calculate the total intensity and fractional polarization from the standing
shock, the emission region electron distribution function is assumed to be a
power law, $dn=K_e E^{-p}dE$, where the spectral index is related to the electron
distribution function by $p=2\alpha+1$.  We assume the jet is unresolved and the
emission is mostly concentrated in a cylindrical shell centered on the local
symmetry axis so that the total intensity and fractional polarization are
\citep{lyut05}:
\begin{align}
I&\approx K(\cos^2{\psi'}+\cos^2{\theta'}-3(\cos{\theta'}\cos{\psi'})^2+1) \label{I} \\
P&\approx\frac{3+3\alpha}{5+3\alpha}\frac{-2(1+3\cos{2\psi'})\sin^2{\theta'}}{5-\cos{2\theta'}-\cos{2\psi'}-3\cos{2\theta'}\cos{2\psi'}},
\label{P}
\end{align}
where $\theta'$ and $\psi'$ are the variables defined in equations (\ref{theta})
and (\ref{psi}) respectively, and $K$ is an arbitrary constant that depends on
emission region details such as the beaming factor, emission region size,
relativistic particle density, and magnetic field strength
\citep{lyut05}.  The sign of $P$ indicates whether the EVPA is parallel
($P>0$) or perpendicular ($P<0$) to the jet frame local magnetic symmetry axis in
the standing shock.  At any time $t'$, the values of $\theta'$ and $\psi'$
represent the particular orientation of the kinked magnetic field and the value
of the magnetic pitch angle in the standing shock respectively.  (See the
appendix for a derivation of equations \ref{I} and \ref{P}.)

\subsection{Polarization and Spectral Behavior of HST-1:  Interpretation}

We believe that the most consistent explanation for the variability observed in
HST-1 is that the flare occurred in a shock within the jet, with the maximum
polarization coming at the time of maximum compression and also maximum optical
flux.  The high polarization ($P$ at maximum in excess of $40\%$) and alignment
of the EVPA with the jet axis in HST-1, rules out the conical reconfinement
hydrodynamic shock model of Nawalejko (2009; see also Bromberg \& Levinson 2009). 
Assuming that the jet magnetic field is weak and tangled, such models cannot produce
a polarization higher than $\sim 30\%$. Furthermore, such a model would 
predict a substantially different orientation of EVPA 
for small and intermediate jet inclinations.  The EVPA we observe in HST-1 is 
very nearly perpendicular to the jet, and nearly constant, so a more consistent
explanation for the data is that the non-thermal activity is restricted to a
localized, perpendicular
(possibly stationary) strong shock within the interior of the flow, as envisioned
in \S 5.1 (see eq. (1)), which is also more
in line with the observed radio morphology.  The feature producing the flare may 
then be associated with the Mach disk produced around the nozzle (converging point)
of the reconfinement shock 
(Stawarz et al. 2006, Paper IV, Gracia et al. 2009).  In the latter model (Gracia
et al. 2009) the synchrotron emissivity is proportional to the comoving frame electron 
density and magnetic field strength and configuration, modulo $\delta^{(2+\alpha)}$, 
rather than being dependent on hydrodynamic energy dissipation as assumed in 
Nawalejko (2009).

Using the doubling/halving timescales calculated for the optical/UV emission  in
Madrid (2009) as well as those in Paper II for the X-ray emission,  we can
estimate a size of $\leq 0.5 \delta$ light-years (where $\delta$ is the Doppler
factor) for the size of the  flaring region,  with the irregularity of the
available measurements perhaps arising because of the complexity of the
compression mechanism (we do not  believe it arises because we did not sample
adequately, given the  smoothness of the observed lightcurve). This is consistent
with the fact that HST-1 is not  resolved by HST, which sets a hard upper limit
of 2 pc on the radius of the optically emitting components (Paper II, Madrid
2009).  Interestingly, however, as already noted in Paper II and Madrid (2009),
the X-ray and optical/UV doubling/halving timescales do not lead to significantly
different constraints for the region size.  

If we then apply equation (1) to knot HST-1, we can reproduce the type of
behavior seen in the $(I, P)$ plane (Figure  3).  But equally interestingly,
we can also constrain the kinematics in the local flow under certain assumptions.
To illustrate, 
we choose $k=0.25$, appropriate for a strong relativistic shock (Meisenheimer et
al. 1989), spectral index  $0.45\leq \alpha\leq 0.55$, 
corresponding to the
mean observed  during both flare ``loops'' (Figures 2, 4),  and 
$0.35 \leq P \leq 0.45$, consistent with 
the peak value attained at both flux maxima.   We then
allow the viewing angle $\theta_{ob}$ and Doppler factor $\delta$ to vary.  We
constrain the solutions so that the apparent speed $\beta_{app}$ falls within the 
range $4.0<\beta_{app}<4.5$, the values reported for VLBI components in Paper IV.
The result is shown in Figure 6.  As can be seen, the permitted values of
$\theta_{ob}$ range between $11-18^\circ$, and $\delta$ can range between 2.5 and
5.5.  The Lorentz factor, $\Gamma$ is more stable, however, with permitted values
ranging between 4.1 and 4.8. These values are in agreement with the upper limit
of $\delta=8$ calculated by Waters \& Zepf (2005) based on other considerations,
and are also consistent with the requirement of
$\theta_{ob} < 19^\circ$ from the  somewhat
faster superluminal speeds seen in HST monitoring during the 1990s 
(Biretta et al. 1999). 
It should be mentioned that with these values of $\alpha$, $P$ and $k$ there are
no allowed solutions with $\Gamma < 2$, and moreover, nearly all the allowed 
solutions with $\Gamma < 3$ require values of $\theta_{ob}>20^\circ$, which is
not allowed.  
\begin{figure}
\includegraphics{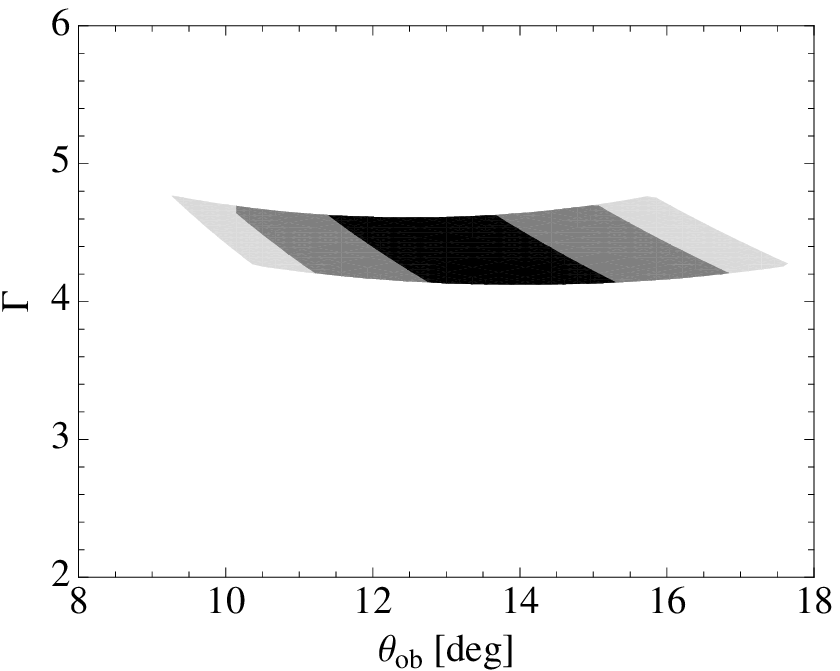}
\includegraphics{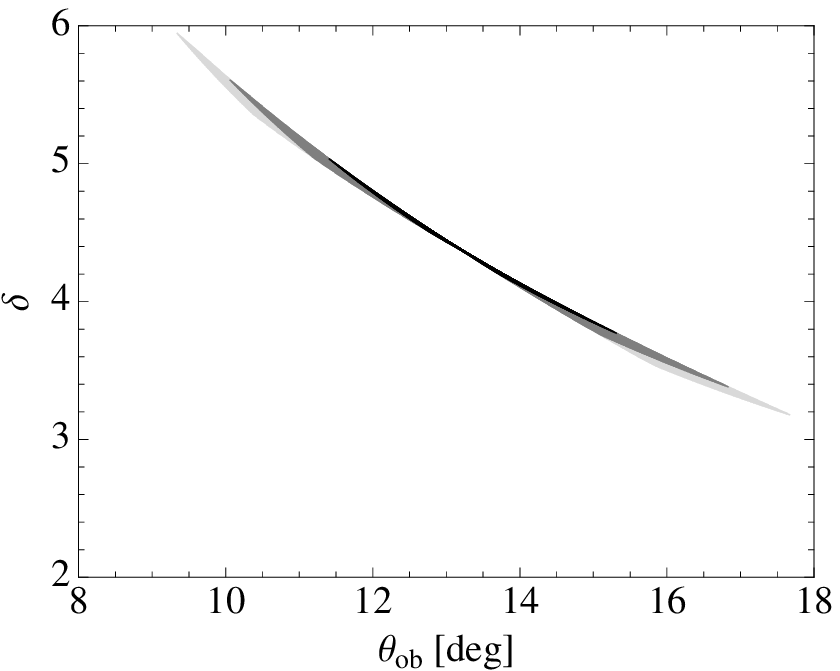}
\caption{Allowed values of beaming parameters for our shock model of knot HST-1.
At top, we show 
the allowed range for the Lorentz factor $\Gamma$ and viewing angle
$\theta_{ob}$, while at bottom, we show the corresponding range for the Doppler
factor $\delta$ plotted against viewing angle $\theta_{ob}$.  In both panels, 
light gray colors refer to spectral indices $\alpha=0.45$, gray colors refer to
$\alpha=0.5$, and black colors refer to $\alpha=0.55$  See \S\S 5.1, 5.2 for
discussion.} 
\end{figure}


Interestingly  however, we see no evidence for motion of the flaring region  of
HST-1 in our data,  as our astrometric results (all epochs having identical
distances between the nucleus and HST-1 to a tolerance of $\pm 0.01''$)   
translate to an upper limit of 1.02 $c$ ($2 \sigma$) on motion of the flaring
region itself.  While this seems to conflict with the VLBA measurements of Paper
IV,  as well as our calculation of $\delta$ from the shock model, this should not
be too concerning. It is entirely possible that the flaring region itself
represents a standing shock, not flowing along with the plasma.  Under such a 
scenario we could not expect the observed speed to accord with the local jet
Lorentz factor.    Indeed, the VLBA maps (Paper IV) show a stationary upstream
end to the HST-1 region.  For most of the time period covered by Paper IV, the
observed speeds are consistent with our limit, as the flaring region (component
HST-1c in their nomenclature), while downstream from the stationary upstream end
(HST-1d), has a speed of  $1.14 \pm 0.14 ~c$ , measured relative to HST-1d. 
Beginning around epoch 2006.0, however, HST-1c splits into two components, with
the faster, downstream one  (accounting for the majority of the radio flux)
accelerating to $4.3 \pm 0.7 ~c$ relative to HST-1d.  While this latter speed is
on the surface highly inconsistent with our astrometric results, it is important
to realize that our data are not very sensitive to this time period,  as it
contains only 5 of our 17 ACS epochs (the lower-resolution  WFPC2 observations of
epoch 18 are much less useful for astrometry).  Using only our ACS data during
2006, the limit on motion of the flaring region would be much less restrictive,
i.e., $\sim 5.5 ~c$ at 2$\sigma$.  Use of later epoch data would improve this
result; however, there was a gap in UV  monitoring of the M87 jet between late
2006 and mid-2009,  a time interval that featured a further factor $\sim 5$
decrease in the X-ray flux of HST-1 (e.g., Paper V and later data), which could
make it very difficult to identify the flaring components. The necessary data do,
however, exist in the VLBA archive, and it would be  highly interesting to track
the motion of the flaring region in both bands.

If indeed the giant HST-1 flare was a result of enhanced particle acceleration
occurring within a compressing shock, then it would be logical to ascribe 
the cooling during the main flare to relaxation of the compression
that occurred within the shock.  In that case, the energy dependence -- or lack
thereof -- of the variability timescales (in optical and other bands) 
may be set by the dynamical
timescale of the compression, as originally noted in Paper II and explored
further in Paper V and Madrid (2009). For example, if the compression and
subsequent expansion were adiabatic, one would expect to see
frequency-independent variability behavior unless there was an intrinsic,
pre-existing break in the spectrum.  This would be modified where the particle 
cooling and acceleration
timescales ($t_{acc}$ and $t_{cool}$ respectively) 
are equal to or less than the dynamical timescales. 
As already noted, we observe hard lags in the optical-UV during the
brightest part of the flare (epochs 5-9, Figure 4).  To explain such hard lags, 
$t_{acc}$ must be similar to $t_{cool}$, whereas when soft lagging was observed
(epochs 13-17), the opposite relationship would hold, i.e., $t_{acc} < t_{cool}$.
Interestingly, as shown in Paper V, in both the UV and X-rays the derivative 
$dI/dt$ changed sign between
2005.4-2005.5, i.e., epochs 9 and 10.  If indeed this was related to the
relationship between the acceleration and cooling timescales, then the
oscillations seen in epochs 10-13 -- which occurred during a time when the 
flux was decreasing monotonically -- become important.  
The multiwavelength spectral characteristics of HST-1 discussed above could be
possibly explained in the framework of the scenario in which the primary loss 
mechanism in the optical-UV is Comptonization of external radiation and the 
optical-UV emitting electrons are near the transition between the Thomson and
Klein-Nishina regimes (see the discussion in \S 6, below).

As discussed in \S 4, the mean EVPA in HST-1 is
somewhat different from the PA of the jet.
Thus while a shock morphology is likely for HST-1, the polarization  data hints
at a more complex morphology than indicated by the simple, unresolved appearance
shown by the images. Two interpretations are possible.  The first possibility is
that HST-1's optical polarized flux comes from two or more  regions.  
While this might seem the simplest interpretation, it is difficult to
reconcile this with the strong correlation between flux and polarization,
combined with the constant EVPA.  In addition, the unresolved nature of the
flaring region in our data and that of Madrid  (2009), constrains the maximum
separation of these components to $\sim 2$ parsecs, which translates to a
constraint on the cooling timescale that, as already discussed,  is consistent
with the spectral evolution we see.  Alternatively, it could indicate either a
twist in the field within the shocked region and/or an oblique shock, as
suggested for the knot A region by Bicknell \& Begelman (1996).  In the latter
case, one might expect to see evidence of a slight local deviation in the flow
direction,  and indeed, the VLBA components seen in
Paper IV do have a significant range of radius vector PAs as measured from the 
standing feature at the upstream end of the HST-1 complex.   Furthermore, there
is weak evidence of small changes in EVPA over time, with the EVPA near the two
flux maxima (epochs 8, 9 and 16) being within 2-3 degrees of the PA of the radial
vector from the nucleus to HST-1 from the VLBA data, while in the other 15 epochs
the EVPA is more closely aligned with the motion vector of the faintest feature 
of the HST-1 complex as seen on the VLBA maps, namely component ``a'', at  the
downstream end.  Note however
that if the shock lies well within the interior of the jet these
deviations might not be indicative of an overall deflection of the flow (as in
knot A).  On a somewhat related note, Nakamura et al. (2010) recently proposed
a model for the overall structure of the M87 jet, in which 
the main shocks in the M87 jet have
a helical magnetic field structure, which could be produced via two methods. 
Bicknell \& Begelman (1996) suggested that the knots in the M87 jet are
generated by the helical modes, which can produce lateral oscillation of the
entire jet but has trouble producing filaments within it.  Hardee \& Eilek
(2011), by contrast,  postulate that the elliptical Kelvin-Helmholtz modes
dominate, and model the  inner M87 jet as having twisted, high-pressure
filaments generated by elliptical Kelvin-Helmholtz instabilities which
eventually disrupt the jet beyond knot A.   Assuming that the magnetic field
within the components is perpendicular to the  local bulk velocity vector
(rather than the mean jet axis vector), such a model can produce the slight
offset between EVPA and jet PA  seen in nearly all of the observations of 
HST-1.
  

\subsection{Polarization Behavior of the Nucleus:  Interpretation}

Because the polarization and flux variability of the nucleus 
forms a coherent pattern--a
counter-clockwise loop in $I$ -- $P$ space, over the course of a year--the
available data challenge scenarios where the variability of the nucleus is
dominated by multiple independent components.  The most natural interpretation of
the nuclear polarization variability is that within the small aperture chosen
(which represented approximately 2 HST resolution elements) the structure of the
M87 jet is relatively simple, perhaps having a single region of the jet
dominating the flux and polarization variability at any one time.  This scenario
is consistent with either models of a magnetically dominated jet in which
large-scale magnetohydrodynamic instabilities play an important role
\citep[e.g.][]{Giannios:2006} or with models of a non-stationary flow through a
standing shock.  
Within such a model there are a number of ways to explain the behavior seen 
in the nucleus.  In \S 5.1 we have illustrated two possible scenarios that may 
occur -- namely, either (1) there is a standing shock through which a time
varying flow propagates, or (2) the jet contains a large-scale helical field 
subject to the kink mode in a region of a standing shock as shown in Figure
\ref{helical_cartoon}.  Both of these
possibilities (see the discussions surrounding equations
(2)-(5) and (6)-(9) respectively), are motivated by (i) the variability 
of jet flow speeds as seen in superluminal studies of jets and (ii) the 
prominence of the kink mode in theoretical analyses of magnetically dominated 
jets respectively.

\begin{figure}
	\centering
		{\includegraphics[width=3.1in]{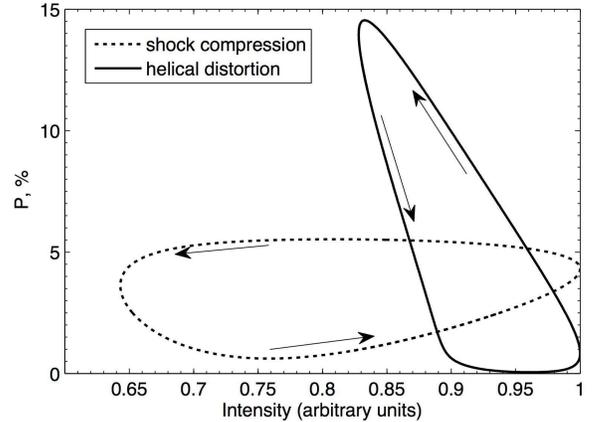}}
		\caption{Illustrated here is a theoretical counter-clockwise $I$
-- $P$ loop for the helical distortion model (solid) and the variable shock
compression model (dotted).  The values used in the shock compression model are:
$(\beta_{u,0}, A_{\beta},\omega_{\beta},\phi_{\beta})=(0.78,0.03,1,-\pi/2)$ and
$(\xi_0, A_{\xi},\omega_{\xi},\phi_{\xi})=(0.78,-0.5,1,0)$.  For the helical
distortion model,
$(A_{\theta}',\omega_{\theta},\phi_{\theta},\theta_{ob}')=(1.5,1,0,\pi/2)$ and
$(A_{\psi}',\omega_{\psi},\phi_{\psi},\psi_0')=(0.05,2,0,1.03)$ and
$\delta=\sin{\theta_{ob}'}/\sin{15^{\circ}}\approx 3.86$.  For both models,
$\theta_{ob}=15^{\circ}$ and $\alpha=1.5$ were used.}
		\label{theory_loop}
\end{figure}

The two scenarios considered here each produce different tracks in the $(I,P)$
plane.  In Figure \ref{theory_loop}, we show an example track for each,
assuming a viewing angle $\theta_{ob}=15^\circ$, consistent with the observation
of superluminal motions at speeds as high as $6c$ in HST-1 (Biretta et al. 1999,
Paper IV; see also \S 5.2).  
As shown, both scenarios successfully reproduce two general
characteristics of the looping behavior we see in the $(I,P)$ plane between 
epochs 6 to 15 in Figure 3, namely: (i) successive
points are located close together in $(I,P)$ space, and (ii) these points form
a coherent pattern (a loop) over the course of a $\sim$ year as shown in figure
\ref{theory_loop}.  Nonetheless, both models suffer from shortcomings. 
Constraining the shock compression model to produce EVPAs parallel to the jet
direction severely restricts the maximum polarization it can achieve. Thus, the
model's maximum polarization reaches only $\sim6\%$, while the observations
achieve a maximum of $\sim 11\%$.  
In the helical distortion model, maintaining a
low polarization and producing an EVPA parallel to the jet implies restricting
the variability of the magnetic pitch angle to be
$\psi'=1/2\cos^{-1}{(-1/3)}+\epsilon$, where $0<\epsilon\ll1$, which is an
arbitrary constraints on the oscillation of $\psi'$.  More generally, both models
are limited by their extreme simplicity (e.g. neither have sheared velocity
fields), high number of free parameters, and non-uniqueness.   

Of the two toy models discussed herein, we would argue that the ($I, P)$ behavior
we observe (Figure 3) is closer to that produced by the helical distortion model,
although again, neither model reproduces the observed behavior perfectly. The
shock compression model requires a roughly linear increase of polarization with
flux in its 'positive' stage, followed by a 'plateau' with polarization nearly
constant at its maximum while the flux decreases.  Neither of these is observed
in the nucleus.  Instead we see very steep increases in polarization during
epochs 6-9 (from 2\% up to 12\%), while the flux changes are much smaller. The
pattern does not exactly reproduce what is seen in the helical distortion model
either, however.  Local maxima in flux are seen at epochs 7 and 12.  Of these,
the former corresponds to a local maximum in $P$ while the latter does not;
indeed, the global maximum in $P$ comes at epoch 9, when the flux is declining. 
In  fact, one could actually argue that a 'plateau' is seen between  epochs 9 and
10, as the two polarization measurement are statistically indistinguishable from
one another while the fluxes differ by 25\%.  It is possible that both types of
variations are seen, somewhat out of phase with one another, but while this is
plausible such a model  would produce competing variations that are difficult to
model without introducing {\it ad hoc} assumptions.

Neither of these models is fully successful in reproducing the EVPA fluctuations
we observe (Figure 1).  The helical distortion model predicts that the EVPA time
evolution should correlate with the polarization and intensity.  We expect this
correlation because the helical distortion flowing through the standing shock
causes the orientation of the magnetic field symmetry axis in the post shock flow
to fluctuate as illustrated in figure \ref{helical_cartoon}; in turn, the
post-shock flow EVPA, intensity, and polarization depend on the fluctuating field
geometry. Unlike the helical distortion model, the varying shock compression
model maintains cylindrical symmetry even as other parameters vary with time. 
Thus, the shock model predicts the EVPA should remain stationary or undergo rapid
$90^{\circ}$ flips.  It should be noted, however, that the dominant observed EVPA
of the nucleus is close to parallel to the jet, an orientation which both toy
models reproduce for the parameters used in figure \ref{theory_loop}.

The lack of variability we observe in $\alpha_{O-UV}$ for
the nucleus ostensibly fits better within the helical distortion than with the 
shock compression model.  In the helical distortion model, the shock strength 
and particle distribution parameters, $K_e$ and $p$ ($=2\alpha+1$), are held 
constant and only the magnetic geometry fluctuates, hence the lack of spectral 
index variation in the model.  However, more realistically, in particle 
acceleration models such as diffusive shock acceleration (or first-order Fermi 
acceleration), the particle acceleration efficiency depends on the angle 
between the large-scale magnetic field and the shock plane (Sironi \& 
Spitkovsky 2011).  In the case of the shock compression model, the increased 
shock compression ratio leads to enhanced particle
acceleration, as envisioned in \S 5.2 for HST-1, the expectation is a strong 
correlation between flux and $\alpha_{UV-O}$, either hard or soft lagging
depending on the relationship between the acceleration and cooling timescales.

\section{Summary}

This work has revealed the value of polarization and spectral information in 
monitoring campaigns, as well as in understanding the physics of jet regions.  We
have shown that knot HST-1 can be best understood as a shocked region, where the
flaring upstream end did not move significantly between 2002-2007, and displays
characteristics that are consistent with a classic perpendicular shock.  
Under such a model we find that the Lorentz factor in the jet at HST-1 can be
constrained to a fairly narrow range, $4.1 \leq \Gamma \leq 4.8$, and the 
viewing angle can also be constrained to $10^\circ \leq \theta_{ob} \leq
18^\circ$.  
The optically emitting part of the jet within HST-1 may well contain multiple
components (as the radio emitting part does, see Paper IV), given the fine
points of the EVPA structure (\S 5.2); however the HST data does not have the
angular resolution to resolve this region.
The spectral behavior of HST-1 was consistent with nearly equal particle
acceleration and cooling timescales in the optical-UV; 
however, the fact that the
X-ray emission is best understood as synchrotron radiation complicates things.  
Two interpretations are possible for this duality.  The first of these is
that the X-ray emission may come from a small part of the jet in HST-1 (as
suggested for downstream regions of the jet by Perlman \& Wilson 2005), which
may either be partially co-spatial with the optical emission region or distinct
from it.  This interpretation is disfavored, however, because of the overall 
similar appearance of the UV and X-ray lightcurves (Paper V).  A more likely
interpretation is that, while the likely emission mechanism in both bands
remains synchrotron radiation, the dominant energy loss mechanism 
is inverse-Comptonization of external radiation, with  
the X-ray emitting electrons being in the
Klein-Nishina regime, while the optical/UV emitting electrons are either in
the Thomson regime or around the Thomson/Klein-Nishina transition.  Under
such a scenario the cooling timescale of lower-energy optical-UV photons may be
comparable to or shorter than the cooling timescale of X-ray emitting electrons. 
The flat optical spectra observed in HST-1 ($\alpha \lsim 0.5$) are in fact
very consistent with such an idea (Moderski et al. 2005).  In such a scenario
the inverse-Compton emission would dominate at higher energies and in fact would
be energetically dominant when integrated over the entire electromagnetic
spectrum.  This would be consistent with the correlation of the 2005 TeV flare
with the HST-1 flare seen at lower energies (Abramowski et al. 2011). 

The
variability behavior seen in the nucleus, however, can be best understood as 
either a helical distortion to a steady jet, where the distortion would arise
from kink mode instabilities, or fluctuations in the jet speed that produce
corresponding fluctuations in the strength of shocks within the nuclear jet. 
Both of these toy models can produce the fluctuations in field components that 
can result in the 'looping' polarization behavior we see. 
We believe the most likely mechanism is a
current-driven instability combined with a fluctuating, helical magnetic
pitch   angle.  Jet precession can be ruled out as there is no evidence of
jet wobbles on larger resolved scales.  This leaves large-scale instabilities as
a possible cause of the wobbling axes.  Kelvin-Helmholtz instabilities could
cause the  wobbling; however these are important in hydrodynamic jets and as
discussed in Section 5.3 we believe that in the nuclear regions the M87 jet is
more likely  to be magnetically dominated, with a helical magnetic field
structure.  The kinked jet scenario naturally leads to magnetic pitch angle
fluctuations.   Future monitoring of the  variability behavior of spatially
resolved jets should include both multi-band imaging and polarimetry on the same
timescale in order to maximize the  physical information that can be gained from
the campaign.

\begin{acknowledgments}

We thank an anonymous referee for thoughtful comments that significantly improved
the quality of this paper.  ESP, MG, MC and RCS acknowledge support from NASA
under LTSA program grant NNX07AM17G and HST grant GO-11138.01. EC 
acknowledges support from the NASA \textit{Fermi} grant NNX10AO46G.
SCA and MB
acknowledge SARA REU fellowships, funded by the National Science Foundation
Research Experiences for Undergraduates (REU) program through grant NSF
AST-1004872.  ESP acknowledges interesting discussions with L. Sironi and M.
Birkinshaw which helped with understanding the variability behavior we see.

\end{acknowledgments}

\appendix
\section{Helical field intensity and polarization}

Here we derive the expressions for synchrotron intensity (eqn. \ref{I}) and
polarization (eqn. \ref{P}) of an unresolved jet with an emitting cylindrical
shell containing a helical field, a configuration first considered in the
nonrelativistic case by \citep{Laing:1981}.  
The notation and discussion here follows that
of \cite{lyut05}, who explicitly derive the polarization expression in
their equation (21).  The synchrotron intensity and polarization can be expressed
as integrals over the entire cylindrical shell:
\begin{align}
I&=K\int^{2\pi}_0{\left|B'\sin{\chi'}\right|^{\alpha+1}d\phi} \\
P&=\frac{3+3\alpha}{5+3\alpha}\frac{\int^{2\pi}_0{\left|B'\sin{\chi'}\right|^{\alpha+1}\cos{2\tilde{\chi}}d\phi}}{\int^{2\pi}_0{\left|B'\sin{\chi'}\right|^{\alpha+1}d\phi}}
\label{IP}
\end{align}
where $\phi$ is the azimuthal cylindrical coordinate, $K$ is a constant, $\chi'$
is the angle between the jet frame line of sight and the jet frame magnetic
field, and $\tilde{\chi}$ is the angle between the EVPA and the jet's projected
onto the sky, measured clockwise.  To obtain analytical results, we assume
$\alpha=1$ in carrying out the above integration.

Assume a Cartesian coordinate system centered on the jet with the bulk flow
directed along the $z$-axis and with the observer in the $y=0$ plane.  Therefore,
the jet velocity is
\begin{align}
\vec{\beta}=\beta(0,0,1)
\end{align}
and the photon propagation vector is
\begin{align}
\hat{n}=\left(\sin{\theta_{ob}},0,\cos{\theta_{ob}}\right)   
\end{align}
The jet frame shell magnetic field unit vector has a pitch angle of
$\tan{\psi'}=B_{\phi}'/B_z'$ and can be expressed in cylindrical coordinates as
\begin{align}
\hat{B}'=\left(-\sin{\psi'}\sin{\phi},\sin{\psi'}\cos{\phi},\cos{\psi'}\right).
\end{align}
Thus $\sin{\chi'}=\vec{n}'\cdot\hat{B}'$ can now be expressed as:
\begin{align}
\sin^2{\chi'}=\cos^2{\psi'}\sin^2{\theta_{ob}'}+\frac{1}{2}\sin{2\theta_{ob}'}\sin{2\psi'}\sin{\phi}+\left(\cos^2{\theta_{ob}'}+\cos^2{\phi}\sin^2{\theta_{ob}}\right)\sin^2{\psi'}.
\label{sin}
\end{align}
Integrating this expression over $\phi$ for $\alpha=1$ produces the intensity
\begin{align}
I=K\left(\cos^2{\psi'}+\cos^2{\theta_{ob}'}-3(\cos{\theta_{ob}'}\cos{\psi'})^2+1\right).
\end{align} 
To find the polarization, all that remains to be calculated is
$\cos{2\tilde{\chi}}$.  First, note that the polarization vector of a synchrotron
electromagnetic wave in the jet frame is $\hat{e}'=\vec{n}'\times \hat{B}'$. 
$\tilde{\chi}$ may now be written as
\begin{align}
\cos{\tilde{\chi}'}=\hat{e}'\cdot\left(\vec{n}'\times \vec{\ell}\right),
\end{align}
where $\vec{n}'\times \vec{\ell}$ is the jet direction projected onto the sky,
and $\vec{\ell}=(0,1,0)$ in Cartesian coordinates.  Evaluating
$\cos{\tilde{\chi}'}$ leads to
\begin{align}
\cos{2\tilde{\chi}'}=\frac{\cos^2{\phi}\sin^2{\psi'}-\left(\cos{\psi'}\sin{\theta_{ob}'}+\cos{\theta_{ob}'}\sin{\phi}\sin{\psi'}\right)^2}{1-\left(\cos{\psi'}\cos{\theta_{ob}'}-\sin{\theta_{ob}'}\sin{\phi}\sin{\psi'}\right)^2}.
\label{cos}
\end{align}
Therefore, according to equation (\ref{IP}),
\begin{align}
P=\frac{3/2\left(1+3\cos{2\psi'}\right)\sin^2{\theta_{ob}'}}{5-\cos{2\theta_{ob}'}-\cos{2\psi'}-3\cos{\theta_{ob}'}\cos{2\psi'}}.
\end{align}
Note that equations (\ref{sin}) and (\ref{cos}) are equations (20) of
\cite{lyut05}. Although this result strictly holds for $\alpha=1$, it can
be extended to other values close to $\alpha=1$ by the following analytic
approximation
\begin{align}
P=\left(\frac{3+3\alpha}{5+3\alpha}\right)\frac{2\left(1+3\cos{2\psi'}\right)\sin^2{\theta_{ob}'}}{5-\cos{2\theta_{ob}'}-\cos{2\psi'}-3\cos{\theta_{ob}'}\cos{2\psi'}}.
\end{align}


\begin{thebibliography}{}

\bibitem[Abdo et al. (2009)]{Fermi} Abdo, A., et al., 2009, ApJ, 707, 55 

\bibitem[Abramowski et al. (2011)]{Raue11} Abramowski, A.,  et al., 2011, ApJ, submitted

\bibitem[Albert et al.(2008)]{2008ApJ...685L..23A} Albert, J., et al.\
2008, \apjl, 685, L23

\bibitem[Acciari et al.(2008)]{2008ApJ...679..397A} Acciari, V.~A., et
al.\ 2008, \apj, 679, 397

\bibitem[Acciari et al.(2009)]{2009Sci...325..444A} Acciari, V.~A., et
al.\ 2009, Science, 325, 444

\bibitem[Acciari et al. (2010)]{Acc10} Acciari, V. A., et al., 2010, ApJ, 716,
819

\bibitem[Aharonian et al. (2003)]{aha03} Aharonian, F., et al., 2003, A\&A,
403, L1

\bibitem[Aharonian et al. (2006)]{aha06} Aharonian, F., et 
al.\ 2006, Science, 314, 1424

\bibitem[Asada et al. (2002)]{Asada:2002} Asada, K., Inoue, M., Uchida, Y., 
Kameno, S., Fujisawa, K., Iguchi, S., Mutoh, M., 2002, \pasj, 54, L39

\bibitem[Bateman (1978)]{Bateman:1978} Bateman, G., 1978, MHD Instabilities,
(Cambridge, MA:  MIT Press)

\bibitem[Begelman (1998)]{Begelman:1998} Begelman, M. C., 1998, ApJ, 493, 291

\bibitem[Bicknell \& Begelman (1996)]{bb96} Bicknell, G. V., Begelman, M. C., 
1996, ApJ, 467, 597

\bibitem[Biretta(1996)]{bir96} Biretta, J. A.\, 1996, The Wide Field and
Planetary Camera 2 Instrument Handbook (Baltimore, MD: STScI)

\bibitem[Biretta \& McMaster (1997)]{mcm97} Biretta, J. A., McMaster, M.\, 1997,
WFPC2 Polarization Calibration Handbook, WFPC2 Instrument Science Report 1997-11
(Baltimore, MD: STScI) 

\bibitem[Biretta et al. (1999)]{Biretta:1999} Biretta, J. A., Sparks, W. B.,
Macchetto, F. D., 1999, ApJ, 520, 621


\bibitem[Bohlin(2007a)]{boh07} Bohlin, R. C.\, 2007a, Instrument Science Report
ACS 2007-06

\bibitem[Bohlin(2007b)]{boh07b} Bohlin, R. C., 2007b, in {\it The Future of
Photometric, Spectrophotometric and Polarimetric Standardization}, ASP Conf.
Series 364, ed. C. Sterken (San Francisco:  PASP), p. 315

\bibitem[B\"ottcher \& Chiang (2002)]{bc02} B\"ottcher, M., Chiang, J., 2002,
ApJ, 581, 127

\bibitem[Bromberg \& Levinson (2009)]{bl09} Bromberg, O., Levinson, A., 2009, 
ApJ, 699, 1274

\bibitem[Cawthorne et al. (1993)]{Cawthorne:1993} Cawthorne, T. V., Wardle, J. F. C.,
Roberts, D. H., Gabuzda, D. C., 1993, \apj, 416, 519

\bibitem[Chen et al. (2011)]{chen11} Chen, Y. J., Zhao, G. Y., Shen, Z. Q., 2011,
MNRAS, in press, arXiv: 1107.0769

\bibitem[Cheung et al.(2007)]{che07} Cheung, C. C., Harris, D. E. Stawarz, L.\
2007, \apjl, 663, L65

\bibitem[Choudhuri \& K\"onigl (1986)]{Choudhuri:1986} Choudhuri, A. R., \&
K\"onigl, A., 1986, ApJ, 310, 96

\bibitem[Clausen-Brown, Lyutikov \& Kharb (2011)]{Clausen-Brown:2011} 
Clausen-Brown, E., Lyutikov, M., \& Kharb, P., 2011, MNRAS, 415, 2081

\bibitem[Curtis (1918)]{cur18} Curtis, H. D., 1918, Pub. Lick Obs., 13, 9

\bibitem[Fiorucci, Ciprini \& Tosti (2004)]{fct04} Fiorucci, M., Ciprini, S.,
Tosti, G., 2004, A\& A, 419, 25 

\bibitem[Fruchter and Sosey(2009)]{fru09} Fruchter, A. \& Sosey, M. et al., 2009,
The MultiDrizzle Handbook version 3.0, (Baltimore, STScI)

\bibitem[Gabuzda, Murray \& Cronin (2004)]{Gabuzda:2004} Gabuzdda, D. C., 
Murray, \'E, \& Cronin, P., 2004, MNRAS, 351, L89

\bibitem[Georganopoulos et al.(2005)]{geo05} Georganopoulos, 
M., Perlman, E.~S., \& Kazanas, D.\ 2005, \apjl, 634, L33 

\bibitem[Giannios \& Spruit (2006)]{Giannios:2006} Giannios, D., \& Spruit, H. ~C.,
2006, \aap, 450, 887

\bibitem[Giannios, Uzdensky \& Begelman (2010)]{gia10} Giannios, D., Uzdensky,
D. A., Begelman, M. C., 2010, MNRAS, 402, 1649

\bibitem[Giannios, Uzdensky \& Begelman (2009)]{gia09}  Giannios, D., Uzdensky,
D. A., Begelman, M. C., 2010, MNRAS, 395, L29


\bibitem[Hardee \& Eilek (2011)]{HE2011} Hardee, P. E., Eilek, J. A., 2011, ApJ,
in press, arXiv:1104.4480

\bibitem[Harris et al.(2003)]{har03} Harris, D. E., Biretta, J. A., Junor, W.,
Perlman, E. S., Sparks, W. B., \& Wilson, A. S.\ 2003, \apjl, 586, L41

\bibitem[Harris et al.(2006)]{har07} Harris, D. E., Cheung, C. C., Biretta, J.
A., Sparks, W. B., Junor, W., Perlman, E. S., Wilson, A. S., 2006, ApJ, 640, 211

\bibitem[Harris et al.(2009)]{har09} Harris, D. E., Cheung, C. C., Stawarz, L.,
Biretta, J. A., Perlman, E. S.\ 2009, \apj, 699, 305

\bibitem[Hook et al. (2000)]{hook00} Hook, R. N., Walsh, J., Pirzkal, N., \&
Freudling, W., 2000, in {\it ASP Conf. Ser. 216, Astronomical Data Analysis \& 
Software Systems IX}, ed.N. Manset, C. Veillet \& D. Crabtree (San Francisco:
ASP), 671

\bibitem[Hughes et al. (1985)]{hu85} Hughes, P. A., Aller, H. D., Aller, M. F., 
1985, ApJ, 298, 301

\bibitem[Kharb et al. (2009)]{Kharb:2009} Kharb, P., et al., 2009, \apj, 
694, 1485

\bibitem[Kirk, Rieger \& Mastichiadis (1998)]{kirk98} Kirk, J. G., Rieger, F.
M., Mastichiadis, A., 1998, A\& A, 333, 452

\bibitem[Kollgaard et al. (1990)]{Kollgaard:1990} Kollgaard, R. I., Wardle, J. F.
C., Roberts, D. H., 1990, AJ, 100, 1057

\bibitem[Kovalev et al. (2007)]{kov07} Kovalev, Y. Y., Lister, M. L., Homan, D.
C., Kellermann, K. I., 2007, ApJ, 668, L27

\bibitem[Junor, Biretta \& Livio (1999)]{Junor:1999} Junor, W., Biretta, J. A.,
Livio, M., 1999, \nat, 401, 892

\bibitem[Laing (1981)]{Laing:1981} Laing, R. A., 1981, ApJ, 248, 67

\bibitem[Laing (1980)]{Laing:1980} Laing, R. A., 1980, MNRAS, 193, 439

\bibitem[Landau \& Lifshitz (1987)]{Landau:1987} Landau, L. D., \& Lifshitz, E.
M., 1987, Fluid Mechanics (Oxford:  Pergamon)

\bibitem[Lind \& Blandford (1985)]{Lind:1985} Lind, K. R., Blandford, R. D., 
1985, \apj, 295, 358

\bibitem[Lister et al. (2009)]{Lister:2009} Lister, M., et al., 2009, \aj,
138, 1874

\bibitem[Ly et al. (2007)]{Ly07} Ly, C., Walker, R. C., Junor, W., 2007, ApJ,
660, 200

\bibitem[Lyutikov, Pariev \& Gabuzda (2005)]{lyut05} Lyutikov, M., Pariev, V. 
I., Gabuzda, D. C., 2005, MNRAS, 360, 869 

\bibitem[Madrid (2009)]{mad09} Madrid, J. M.\ 2009, \aj, 137, 3864

\bibitem[Marscher (2009)]{Marscher:2009} Marscher, A. P., 2009, in The Jet 
Paradigm: From Microquasars to Quasars, T. Belloni (ed.), Lect. Notes Phys., 794,
arXiv: 0909.2576

\bibitem[Maybhate et al. (2010)]{May10} Maybhate, A., et al., 2010, ACS
Instrument Handbook 

\bibitem[McKinney (2006)]{McKinney:2006} McKinney, J.~C., 2006, MNRAS, 368, 1561

\bibitem[McKinney \& Blandford (2009)]{McKinney:2009} McKinney, J.~C., Blandford, R.~D., 
2009, MNRAS, 394, L126  

\bibitem[Meisenheimer et al. (1989)]{Meisenheimer: 1989} Meisenheimer, K.,
R\"oser, H.--J., Hiltner, P. R., Yates, M. G., Longair, M. S., Chini, R., Perley,
R. A., 1989, A\& A, 219, 63

\bibitem[Moderski et al. (2005)]{Moderski:  2005} Moderski, R., Sikora, M., 
Coppi, P. S., Aharonian, F., 2005, MNRAS, 363, 954

\bibitem[Mizuno et al. (2011)]{Mizuno:2011} Mizuno, Y., Hardee, P. E., Nishikawa,
K. I., 2011, \apj, 734, 19

\bibitem[Mizuno et al. (2009)]{Mizuno:2009} Mizuno, Y., Lyubarsky, Y., Nishikawa,
K.~I., \& Hardee, P. E., 2009, \apj, 700, 684

\bibitem[Mobasher et al.(2002)]{mob02} Mobasher, B., et al.\ 2002, HST Wide
Field and Planetary Camera 2 Data Handbook (Baltimore, MD: STScI)

\bibitem[Naghizadeh-Kouei  \& Clarke (1993)]{nkc93} Naghizadeh-Khouei, J., \& 
Clarke, D., 1993, A\& A, 274, 968

\bibitem[Nakamura, Garofalo \& Meier (2010)]{naka10} Nakamura, M., Garofalo,
D., Meier, D. L., 2010, ApJ, 721, 1783

\bibitem[Nakamura \& Meier (2004)]{Nakamura:2004} Nakamura, M., \& Meier, D. L., 
2004, \apj, 617, 123

\bibitem[Nakamura et al. (2007)]{Nakamura:2007} Nakamura, M., Li, H., Li, S.--T.,
2007, \apj, 656, 721  

\bibitem[Nawalejko (2009)]{naw09} Nawalejko, K., 2009, MNRAS, 395, 524

\bibitem[Nalewajko et al. (2010)]{naw10} Nalewajko, K., Giannios, D., 
Begelman, M. C., Uzdensky, D. A., Sikora, M., 2010, MNRAS, in press,
arXiv:1007.3994

\bibitem[Narayan et al. (2009)]{Narayan:2009} Narayan, R., Li, J., Tchekhovskoy, A.,
2009, \apj, 697, 1681


\bibitem[Owen, Hardee \& Cornwall (1989)]{Owen:1989} Owen, F. N., Hardee, P. E., 
Cornwall, T. J., 1989, \apj, 340, 698

\bibitem[Perlman et al. (2001)]{per01} Perlman, E. S., Biretta, J. A., Sparks,
W. B., Macchetto, F. D., Leahy, J. P., 2001, \apj, 551, 206

\bibitem[Perlman et al.(1999)]{per99} Perlman, E. S., Biretta, J. A., Zhou, F.,
Sparks, W. B., \& Macchetto, F. D.\ 1999, \aj, 117, 2185

\bibitem[Perlman et al.(2003)]{per03} Perlman, E. S., Harris, D. E., Biretta, J.
A., Sparks, W. B., Macchetto, F. D.\ 2003, \apjl, 599, L65

\bibitem[Perlman et al. (2006)]{per06} Perlman, E. S., et al., 2006, ApJ, 651,
735 

\bibitem[Perlman \& Wilson (2005)]{PW06} Perlman, E. S., Wilson, A. S., 2005,
ApJ, 627, 140

\bibitem[Serkowski (1962)]{ser62} Serkowski, K., 1962, Adv. Astron. Astrophys., 
1, 289

\bibitem[Sirianni et al. (2005)]{sir05} Sirianni, M., et al., 2005, PASP, 117,
1049

\bibitem[Sironi \& Spitkovsky (2011)]{Sironi:2011} Sironi, L., Spitkovsky, A.,
2011, \apj, 726, 75

\bibitem[Spitkovsky (2008)]{Spitkovsky: 2008} Spitkovsky, A., 2008, \apjl, 
682, L5

\bibitem[Spruit (2010)]{Spruit:2010} Spruit, H.~C., , in The Jet 
Paradigm: From Microquasars to Quasars, T. Belloni (ed.), Lect. Notes Phys., 794,
233; arXiv:0804.3096

\bibitem[Spruit, Foglizzo \& Stehle (1997)]{Spruit:1997} Spruit, H. C., 
Foglizzo, T., Stehle, R., 1997, \mnras, 288, 333

\bibitem[Stawarz et al. (2006)]{sta06} Stawarz, L., Aharonian, F., Kataoka, J.,
Ostrowski, M., Siemignowska, A., Sikora, M., 2006, MNRAS, 370, 981 

\bibitem[Taylor (1974)]{Taylor:1974} Taylor, J. B., 1974, \prl, 33, 1139

\bibitem[Tomimatsu et al. (2001)]{Tomimatsu:2001} Tomimatsu, A., Matsuoka, T.,
Takahashi, M., 2001, \prd, 64, 123003

\bibitem[Tonry (1991)]{ton91} Tonry, J. L., 1991, ApJ, 373, L1 

\bibitem[Taylor \& Zavala (2010)]{Taylor:2010} Taylor, G. B., Zavala, R., 2010,
\apjl, 722, L183

\bibitem[Tsvetanov et al(1998)]{tsv98} Tsvetanov, Z. I., Hartig, G. F., Ford, H. C., Dopita, M. A., Kriss, G. A., Pei, Y. C., Dressel, L. L., \& Harms, R. J.\ 1998, \apjl, 493, L83

\bibitem[Vlahakis \& K\"onigl (2004)]{Vlahakis:2004} Vlahakis, N. \& K\"onigl, 
A., 2004, \apj, 605, 656

\bibitem[Walker et al. (2009)]{wrc09} Walker, R. C., Ly, C., Junor, W., Hardee,
P., in {\it Approaching Micro-Arcsecond With VSOP-2:  Astrophysics \&
Technologies}, ASP Conf. Series 402, ed. Y. Hagiwara, E. Fomalont, M. Tsuboi, Y.
Murata (San Francisco:  ASP), 227, arXiv:0803.1837  

\bibitem[Wagner et al. (2009)]{wag09} Wagner, R. M., et al., 2009, 
arXiv:0907.1465

\bibitem[Wardle et al. (1994)]{Wardle:1994} Wardle, J. F. C., Cawthorne, T. V.,
Roberts, D. W., Brown, L. F., 1994, \apj, 437, 122  

\bibitem[Wardle \& Kronberg (1974)]{wk74} Wardle, J. F. C., Kronberg, P. P.,
1974, ApJ, 194, 249

\bibitem[Waters \& Zepf (2005)]{wz05} Waters, C., Zepf, S., 2005, ApJ, 624, 656

\bibitem[Zavala \& Taylor (2005)]{Zavala:2005} Zavala, R. T., \& Taylor, G. B.,
2005, \apjl, 626, L73

\bibitem[Zhang (2002)]{zha02b} Zhang, Y.--H., 2002, MNRAS, 337, 609

\bibitem[Zhang et al. (2002)]{zha02} Zhang, Y.--H., et al., 2002, ApJ, 572, 762

\end{thebibliography}
\end{document}